\documentclass[fleqn,usenatbib]{mnras}

\usepackage{newtxtext,newtxmath}

\usepackage[T1]{fontenc}
\usepackage{ae,aecompl}

\usepackage{graphicx}	
\usepackage{amsmath}	
\usepackage{amssymb}	
\usepackage{array}
\usepackage{multirow}
\usepackage{tabularx}

\usepackage[dvipsnames]{xcolor}
\colorlet{orange}{green!10!orange!90!}

\usepackage[normalem]{ulem} 

\title[Clouds vs. Environment]{Cloud populations versus galactic environment in NGC\,4689, NGC\,628, NGC\,1566 and NGC\,4321}

\author[H. Faustino Vieira et al.]{Helena Faustino Vieira$^{1,2}$\thanks{E-mail: \href{mailto:helena.faustinovieira@astro.su.se}{helena.faustinovieira@astro.su.se}},
Ana Duarte-Cabral$^{1}$, Matthew W. L. Smith$^{1}$, Dario Colombo$^{3}$, Timothy A. Davis$^{1}$,
\newauthor
Zein Bazzi$^{3}$ 
\\
$^{1}$ Cardiff Hub for Astrophysics Research and Technology (CHART), School of Physics \& Astronomy, Cardiff University, The Parade, CF24 3AA Cardiff, UK\\
$^{2}$ Department of Astronomy, Oskar Klein Center, Stockholm University, AlbaNova University Center, SE-106 91 Stockholm, Sweden\\
$^{3}$ Argelander-Institut für Astronomie, Universität Bonn, Auf dem Hügel 71, 53121 Bonn, Germany\\
}

\date{Accepted XXX. Received YYY; in original form ZZZ}

\pubyear{2025}

\begin{document}
\label{firstpage}
\pagerange{\pageref{firstpage}--\pageref{lastpage}}
\maketitle

\begin{abstract}

The study of molecular clouds in galaxies beyond the Local Group is limited by the need to efficiently sample diverse galactic environments across galactic discs, typically resulting in a loss of resolution. Using a high-resolution dust extinction technique, we image the dust (and gas) of 4 nearby galaxies (<18\,Mpc; NGC\,4689, NGC\,628, NGC\,1566, and NGC\,4321) with resolutions between 5--9~pc. We present catalogues of spatially-resolved clouds for these galaxies, with which we investigate whether different galactic environments and morphologies have a significant impact on observed cloud properties. We find no systematic differences in cloud size, aspect ratio, or morphology with galactic environment or radius. We do find changes in cloud masses/surface densities between the centres and discs of galaxies, with clouds in centres typically displaying higher values of mass/surface density. Furthermore, we find distinct distributions of cloud surface densities across the bars of NGC\,1566 and NGC\,4321. Differences between the arm and inter-arm populations are more subtle, with some galaxies in the sample having much higher cloud masses/surface densities in their spiral arms, and other galaxies showing fairly similar arm/inter-arm distributions. These results suggest that, even within this small sample of galaxies, not all spiral arms and bars seem to behave and affect the interstellar medium equally. Therefore, performing a qualitative environment analysis, where clouds of different galaxies are all binned together under the same visual environmental classification, leads to the loss of information on interesting property variations which in turn demonstrate the impact of the underlying dynamics.


\end{abstract}




\section{Introduction}

Galaxies in the local Universe come in many shapes and sizes. It is estimated that two out of three galaxies have a spiral structure \citep[e.g.][]{willett_2013,buta_2015}, and that $\sim30$\,--\,60\% display stellar bars in their centre \citep[e.g.][]{de-Vaucouleurs_1991,sheth_2008,masters_2011}. It is therefore important to investigate the impact these morphological features (i.e. the underlying dynamics) might have on the interstellar medium (ISM) and subsequently on star formation since these structures are key drivers of secular evolution in galaxies \citep[for a review, see][]{kormendy_kennicutt_2004}. For example, spiral arms are notorious for accumulating gas, leading to higher star formation rates (whether this is a natural consequence of more material or a "triggering" effect has not yet been determined; e.g. \citealt{foyle_2010,querejeta_2024, sun_2024}), whilst bars can be very efficient at funnelling gas inwards, potentially enhancing star formation and black hole accretion in the centres of galaxies \citep[e.g.][]{sheth_2005,hogarth_2024}. 

Trying to determine if star formation is directly affected by the large-scale dynamics within a galaxy is a matter of long-standing debate. This investigation is hindered by the difficulty of simultaneously probing the small scales associated with the star formation process and the large-scale structures/dynamics that might regulate the formation and evolution of molecular clouds, where stars form. A popular methodology is to statistically analyse molecular cloud populations in search for any systematic trends of cloud properties with galactic environment \citep[e.g.][]{colombo_2014,duarte-cabral_2021,rosolowsky_2021,faustinovieira_2024,sunyan_2024}. This type of analysis is, of course, limited by the achievable physical resolution, and thus this type of work is often restricted to the Milky Way and nearby galaxies ($<20$~Mpc). Recently, from their analysis of cloud-scale ($\sim100$~pc) sightlines across 70 nearby galaxies using CO(2-1), \cite{sun_2020} found high surface densities in galactic centres (particularly in barred galaxies), as well as a moderate increase in surface density in spiral arms relative to inter-arm regions. \cite{querejeta_2021} derive similar conclusions, and the authors do not find a significant increase of star formation efficiency (i.e. star formation rate normalised by amount of available molecular gas) towards spiral arms \citep[see also][]{querejeta_2024}, although they report smaller depletion times on average in galactic centres. In this study, we use an independent approach that is able to probe smaller spatial scales ($\sim10$~pc) to corroborate these findings.

With the goal of generating catalogues of highly-resolved molecular clouds, this paper describes the application of the extinction imaging technique presented in \cite{faustinovieira_2023} to a wider sample of nearby galaxies of different morphology types \citep[see also][]{faustinovieira_2024}. This extinction technique utilises the dust extinction observed at optical wavelengths to produce high-resolution maps of the dust (and gas) of nearby galaxies. In this work, we use extinction-derived maps of 4 nearby galaxies (NGC\,4689, NGC\,628/M74, NGC\,1566 and NGC\,4321/M100) to investigate the distribution of spatially resolved molecular clouds, as well as any trends of cloud properties with the galactic environment. Section~\ref{sec:data} describes the galaxy sample and data products used in this endeavour, and also holds a summary of the extinction technique used here. A brief description of the extraction of resolved cloud catalogues and the derived cloud properties can be found in Section~\ref{sec:cloud_extraction}. In Section~\ref{sec:nat_cat}, we explore trends between a given cloud property and its host galaxy in terms of large-scale environment and galactocentric radius. Section~\ref{sec:uni_cat} investigates correlations between large-scale environments and cloud characteristics at a common, homogeneous resolution across the sample of galaxies. A summary and discussion of the findings can be found in Section~\ref{sec:sum_conc}.

\section{Data}
\label{sec:data} 

In this paper, we apply the extinction technique outlined in \cite{faustinovieira_2023} to NGC\,4689, NGC\,628, NGC\,1566, and NGC\,4321. This technique allows the imaging of the dust (and gas) content in galaxies at high spatial resolution by measuring dust extinction in the optical. It also uses dust emission observations in the far-infrared (FIR) as a benchmark for dust mass estimates. The estimated dust (and gas) surface density maps have $0.11"$ angular resolution, ranging from $\sim5-9$\,pc in physical resolution at the respective distances of the galaxies in the sample. The properties of these nearby disc galaxies are summarised in Table~\ref{tab:gal_properties}. These specific galaxies were selected because of their nearly face-on orientation (which facilitates the application of the extinction technique, given the assumed dust-stars geometry, presented in \citealt{faustinovieira_2023}, as well as the reduced projection effects at low inclinations), the wealth of multi-wavelength data available for these targets, and their morphological type. Indeed, given that the goal of this work is to investigate variations of the ISM as a function of environment, we aimed for a varied morphological sample, selecting a flocculent galaxy (NGC\,4689), a non-barred spiral (NGC\,628) and two barred galaxies of different bar strengths (NGC\,1566 and NGC\,4321).

\begin{table*}
    \centering
    \begin{tabular}{l c c c c c c c c c}
        \hline
        \hline
        Galaxy & $\phi^\mathrm{a}$ & $i^\mathrm{a}$ & $D$ & Morph. & log $M_*$ & $l_\mathrm{ext}$ & $M_\mathrm{ext}^\mathrm{dust}$ & $M_\mathrm{lit, OH94}^\mathrm{dust}$ & $M_\mathrm{lit}^\mathrm{dust}$ \\
         & (deg) & (deg) & (Mpc) & Type & ($\mathrm{M}_\odot$) & (pc) & ($10^6~\mathrm{M}_{\odot}$) & ($10^6~\mathrm{M}_{\odot}$) & ($10^6~\mathrm{M}_{\odot}$)  \\
        (1) & (2) & (3) & (4) & (5) & (6) & (7) & (8) & (9) & (10) \\
        \hline

        NGC\,4689 & 164.1 & 38.7 & 15.0 ($\pm 2.25$)$^\mathrm{b}$ & SA(rs)bc & 10.24 & 8.0 & 3.26 ($\pm 0.1$) & $2.7\,(0.3)$ & $9.0\,(\pm1.1)^\mathrm{c}$ \\
        
        NGC\,628 & 20.7 & 8.9 & 9.84 ($\pm 0.63$)$^\mathrm{d}$ & SA(s)c & 
10.34 & 5.2 & 2.95 ($\pm 0.3$) & $5.3\,(\pm0.7)$ & $29\,(\pm4)^\mathrm{e}$ \\
        
        NGC\,1566 & 214.7 & 29.6 & 17.69 ($\pm 2$)$^\mathrm{f}$ & SAB(s)bc & 10.79 & 9.4 & 10.8 ($\pm 0.1$) & $1.3\,(\pm0.2)$--23 & $4.3\,(\pm0.6)^\mathrm{c}$--$160^\mathrm{g}$\\
        
        NGC\,4321 & 156.2 & 38.5 & 15.21 ($\pm 0.49$)$^\mathrm{h}$ & SAB(s)bc & 10.75 & 8.1 & 11.5 ($\pm 0.3$)& $15\,(\pm2)$ & $51\,(\pm5)^\mathrm{c}$ \\
        \hline
         
    \end{tabular}
    \caption{Summary table of galaxy properties and observational parameters. (1) Galaxy name. (2) Position angle of galaxy. (3) Inclination of galaxy. (4) Distance to galaxy. (5) Morphological type of the galaxy, from the \protect\href{https://ned.ipac.caltech.edu/byname}{NASA Extragalactic Database}, based on \protect\cite{de-Vaucouleurs_1991}. (6) Galaxy stellar mass from \protect\citet{leroy_2021}. (7) Physical linear resolution of extinction-derived gas surface density map of the galaxy. (8) Total dust mass retrieved from extinction-derived surface density map, with the associated uncertainty (see Appendix~\ref{appA:monte_carlo}). (9) Dust mass values from the literature, adjusted to our adopted dust absorption coefficient (see text). (10) Dust mass values from the literature (not adjusted). \\ \textbf{References}: $^\mathrm{a}$~\protect\citet{lang_2020}. $^\mathrm{b}$~\protect\citet{kourkchi_2020}. $^\mathrm{c}$~\protect\citet{nersesian_2019}. $^\mathrm{d}$~\protect\citet{jacobs_2009}. $^\mathrm{e}$~\protect\citet{aniano_2012}. $^\mathrm{f}$~\protect\citet{kourkchi_tully_2017}. $^\mathrm{g}$~\protect\citet{wiebe_2009}. $^\mathrm{h}$~\protect\citet{freedman_2001}.}
    \label{tab:gal_properties}
\end{table*}

\begin{figure*}
    \centering
    \includegraphics[width=0.9\textwidth]{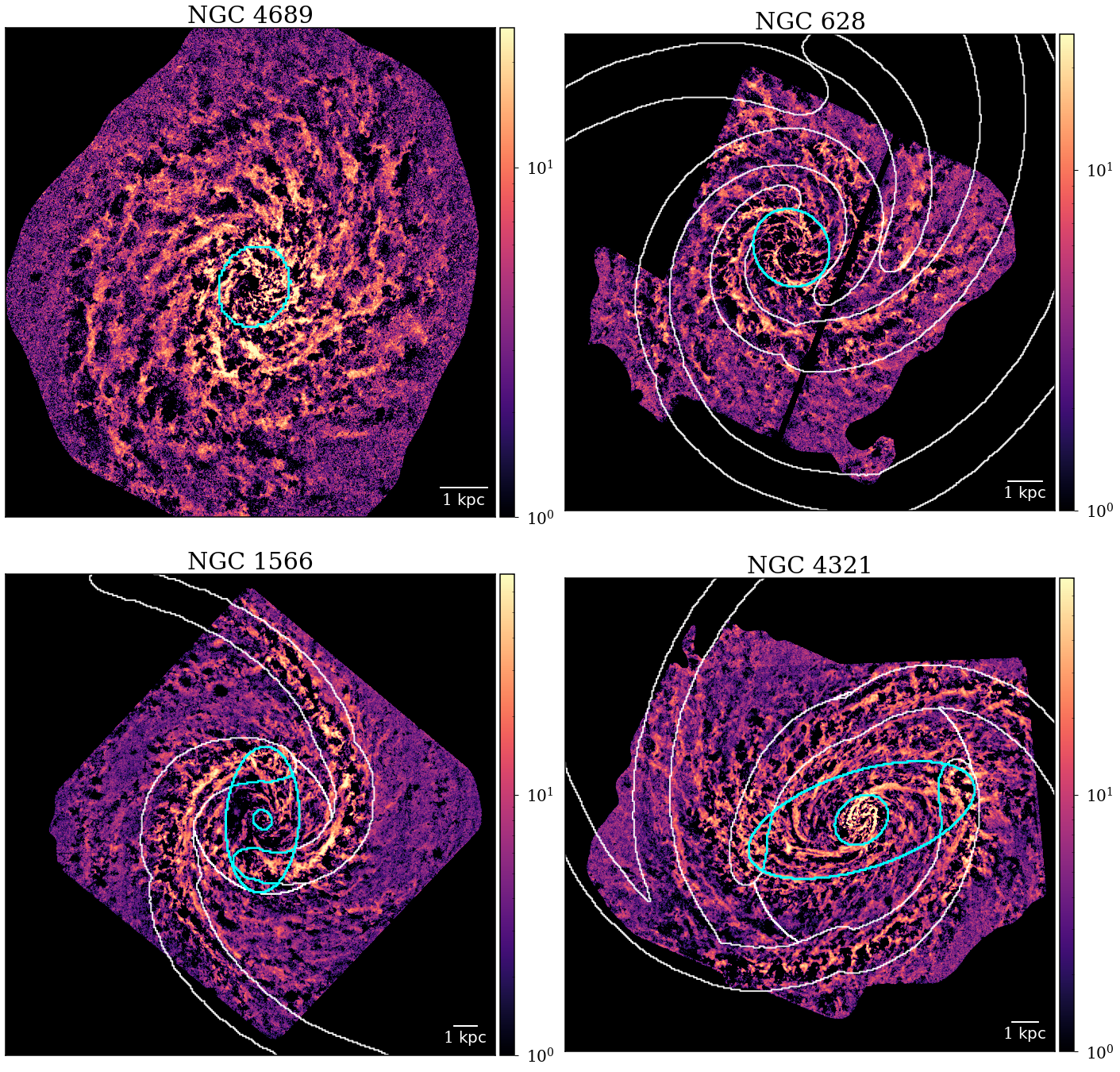}
    \caption{Gas mass surface density maps of the galaxies in the sample (NGC\,4689, NGC\,628, NGC\,1566, and NGC\,4321), in $\mathrm{M}_\odot\,\mathrm{pc}^{-2}$ units. The white contours correspond to the environment masks present in the masks from \citet{querejeta_2021}, which we use in this paper, with the centres/bars highlighted in cyan. Note that the colour-scale ranges are different for the barred and non-barred galaxies.}
    \label{fig:sd_sample}
\end{figure*}

\subsection{Optical data}

For all galaxies, the F555W filter (V-band, centred at 536\,nm) from the \textit{Hubble} Space Telescope (HST) Wide Field Camera 3 (WFC3) was used to build the extinction-derived surface density maps. The optical data for NGC\,1566 is taken from the HST LEGUS (Legacy ExtraGalactic Ultraviolet Survey, \citealt{calzetti_2015}) program (ID13364), whilst the products for NGC\,628\footnote{A small offset in intensity of 0.046~$\mathrm{e^-/s}$ was found between the PHANGS-HST V-band for this galaxy and the earlier dataset towards the same target from LEGUS, due to differences in the calibration used. The PHANGS-HST data was scaled up by the offset, to make the data products consistent with each other.}, NGC\,4321 and NGC\,4689 are taken from the PHANGS-HST (Physics at High Angular Resolution in Nearby Galaxies, \citealt{lee_2022}) program (ID15654)\footnote{\url{https://archive.stsci.edu/hlsp/phangs.html}}. The pixel scale across the sample is $\sim0.04^{\prime\prime}/pix$, corresponding to an angular resolution of $0.08^{\prime\prime}$ (since the PSF is undersampled).

\subsection{Far-Infrared data}
\label{sec:ppmap}

The high-resolution dust extinction imaging technique outlined in \cite{faustinovieira_2023} utilises FIR dust emission observations to ensure the mass estimates from dust extinction are consistent with those from dust emission. The FIR images used here were observed with the \textit{Herschel} Space Observatory \citep{pilbratt_2010}, using both the PACS \citep{poglitsch_2010} and SPIRE \citep{griffin_2010} instruments. All of our targets have observations in the SPIRE bands (250, 350, and $500~\upmu$m) and all three PACS bands (70, 100, $160~\upmu$m), with the exception of NGC\,4689 which was not observed at $70~\upmu$m. The FIR images used in this work have been processed with the pipeline \citep{clark_2018} developed for the DustPedia database \citep{davies_2017}. We recover dust surface densities for each galaxy by modelling the FIR SED with PPMAP \citep[][]{marsh_ppmap_2015}, which is described in Appendix~\ref{app:improvements-SED}.

\subsection{High-resolution extinction technique}
\label{sec:technique}

This paper makes use of the technique presented in \cite{faustinovieira_2023}, and we refer the reader to that paper for the full details. We have implemented several improvements to the technique which are discussed in Appendix~\ref{app:improvements}. As a quick summary, our HST-based extinction technique compares the observed optical light of a galaxy to a reconstructed stellar light model on a pixel-per-pixel basis, which mimics what the galaxy would look like if no dust extinction had occurred. This stellar distribution model is created through median filtering techniques as well as interpolation of the V-band data, first treated to remove bright point-like sources. Assuming that the dust is sitting in a layer close to the mid-plane of a galaxy, in a "sandwich"-like geometry, we can retrieve an estimate for the optical depth, $\tau$, through:

\begin{equation}
    \label{eqn:tau_form}
    \tau = -\mathrm{ln} \left( \frac{I - I_\mathrm{fg}}{I_\mathrm{bg}} \right),
\end{equation}

\noindent where $I$ is the V-band intensity (with bright sources removed). Given our assumed dust/stars geometry, we denote the fractions of the reconstructed stellar light model sitting above and below the dust as $I_\mathrm{fg}$ and $I_\mathrm{bg}$, respectively. This technique includes a calibration step, which ensures that the dust mass estimates from extinction are consistent with those derived from dust emission in the FIR. This is done by adjusting the global background/foreground fraction until the extinction dust masses match those from FIR emission. As mentioned in Section~\ref{sec:ppmap}, we use PPMAP to retrieve an estimate of dust masses from FIR emission, adopting a dust mass absorption coefficient in the infrared from OH94.

It is possible to convert from a measured optical depth to a gas mass surface density ($\Sigma$),

\begin{equation}
    \Sigma =  \frac{\tau}{\delta_\mathrm{DGR} \, \kappa},
\end{equation}

\noindent if we know the dust mass absorption coefficient, $\kappa$, and assume a dust-to-gas mass ratio, $\delta_\mathrm{DGR}$. As in \cite{faustinovieira_2023}, here we adopt a constant $\delta_\mathrm{DGR}=0.01$, and the gas mass absorption coefficient $\kappa_{\mathrm{V}}=1.79$~pc$^2$M$_\odot^{-1}$ (or $\kappa_{\mathrm{V}}=8.55\times10^3~$cm$^2$g$^{-1}$) for the V-band from \cite{draine_2003}. For the FIR regime, we adopt the absorption coefficient {$\kappa_{250\mu\mathrm{m}}=21.6$~$\mathrm{cm}^2\mathrm{g}^{-1}$} at $250~\upmu$m, from \cite{ossenkopf_dust_1994}, with a fixed dust spectral index $\beta$ of 2. In \cite{faustinovieira_2023} we explored the effects of assuming different values of $\kappa_\lambda$, concluding that although the final surface density maps might increase/decrease by a scaling factor, the observed structural hierarchy and environmental trends will not be affected.

Figure~\ref{fig:sd_sample} showcases the final gas mass surface density maps for the 4 galaxies in the sample, which all have an angular resolution of $0.11^{\prime\prime}$ (slightly lower resolution than the original HST resolution, $0.08^{\prime\prime}$, due to a small convolution step in the point-like source removal process, see \citealt{faustinovieira_2023}). Figure~\ref{fig:sd_sample} also shows the environment masks from \cite{querejeta_2021} overlaid as white and cyan contours. To construct these masks, the authors identify galactic environments from near-infrared images (\textit{Spitzer} 3.6~$\upmu$m), which trace the stellar structures of galaxies \citep[see][for details]{querejeta_2021}. Throughout this work we use these masks to investigate systematic trends in our cloud populations. They are composed of 3 major environments: the centre(/bar), C(/B); the spiral arms, SA; and the inter-arm regions, IA. The "centre" environment masks trace the bulge-like, centrally concentrated stellar structure of these galaxies \citep[for further details see][]{querejeta_2021}. For clarity, in this work we use "centre" when denoting this bulge-like central environment for all galaxies in the sample, whilst "centre/bar" is used when referring to the combined central environments of NGC\,1566 and NGC\,4321 (i.e. considering both the centre and bar). NGC\,4689, being a flocculent galaxy, has no well-defined spiral arms, and so we simply denote the entire disc, D.

\subsubsection{Total dust mass estimates}

Table~\ref{tab:gal_properties} lists the extinction-derived total dust mass we retrieve for each galaxy in our sample, alongside the associated uncertainty. These uncertainties are estimated by propagating the Monte Carlo simulations performed for each pixel in our opacity maps (see \citealt{faustinovieira_2023} and Appendix~\ref{appA:monte_carlo} for further details). Also in Table~\ref{tab:gal_properties} are dust mass values from the literature for these galaxies. When comparing dust mass values, it is important to first correct for the different dust model assumptions, which can be done by applying a factor that accounts for the difference in assumed FIR $\kappa$ \citep[see][]{faustinovieira_2023}: 

\begin{equation}
    M_\mathrm{lit, OH94} = \left( \frac{\kappa^\mathrm{lit}}{21.6~\mathrm{cm}^2\mathrm{g}^{-1}} \right) M_\mathrm{lit},
    \label{eqn:adjust_mass}
\end{equation} 

\noindent where $M_\mathrm{lit}$ is a given dust mass from the literature, and $M_\mathrm{lit, OH94}$ the same value but adjusted to our adopted opacity law from \cite{ossenkopf_dust_1994}. The literature values quoted here differ in methodology as well as assumed dust model\footnote{\cite{nersesian_2019} employ both the \texttt{THEMIS} \citep[${\kappa=6.40~\mathrm{cm}^2\mathrm{g}^{-1}}$ at $250\upmu$m,][]{jones_themis_2013} and the \cite{draine_li_2007} (${\sim\kappa=3.98~\mathrm{cm}^2\mathrm{g}^{-1}}$ at $250\upmu$m) dust models. \cite{aniano_2012} also adopts the \cite{draine_li_2007} model, whilst \cite{wiebe_2009} utilises ${\kappa=3.05~\mathrm{cm}^2\mathrm{g}^{-1}}$ at $250\,\upmu$m. Methodology ranges from modified blackbody fits, to \texttt{CIGALE} \citep{noll_2009} spectral modelling, and with a spectral dust index range of $\beta=1.8$--2.}. From Table~\ref{tab:gal_properties} is is possible to see that our dust mass estimates are generally consistent with the literature. Notably, our estimated masses for NGC\,628 and NGC\,4321 reflect the limited FoV of the HST observations compared to \textit{Herschel}, and thus our values are typically lower.

\begin{figure*}
    \centering
    \includegraphics[width=\textwidth]{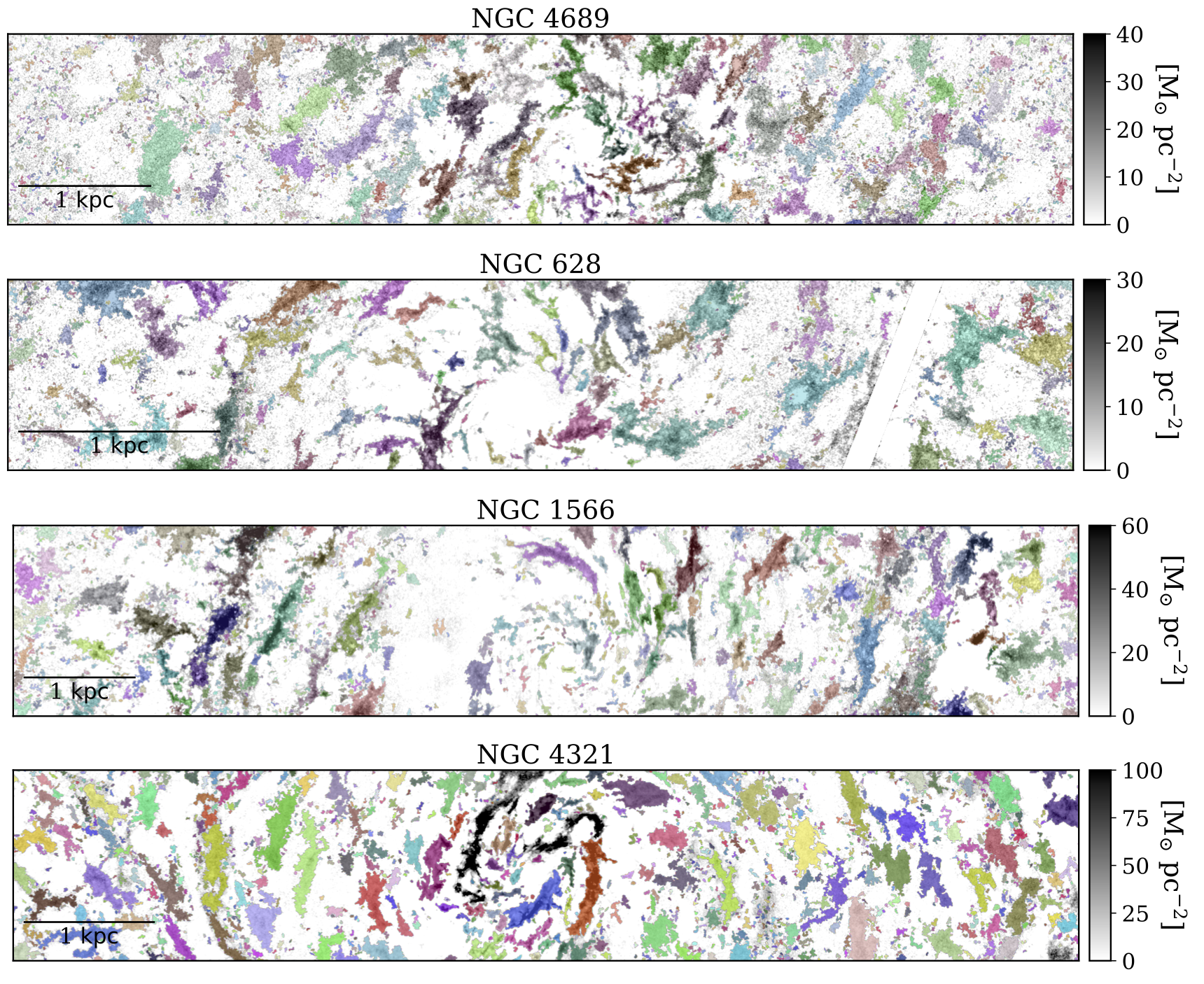}
    \vspace{-0.5cm}
    \caption{Cutout strips across the galactic centre of NGC\,4689, NGC\,628, NGC\,1566 and NGC\,4321 (from top to bottom). Clouds from the native resolution catalogue of each galaxy are plotted in random colours. The gas surface densities are the background greyscale for each galaxy.}
    \label{fig:science_asgn}
\end{figure*}

\section{Cloud populations from HST extinction}
\label{sec:cloud_extraction}

With the goal of statistically studying the variations of ISM properties as a function of large-scale environment, we segment our extinction-derived gas maps into clouds. This is achieved with the use of the \texttt{ASTRODENDRO}\footnote{\href{https://github.com/dendrograms/astrodendro}{https://github.com/dendrograms/astrodendro}} \citep{rosolowsky_2008} package with which we construct the dendrogram of our gas surface density maps, and the spectral clustering algorithm \texttt{SCIMES}\footnote{\href{https://github.com/Astroua/SCIMES/}{https://github.com/Astroua/SCIMES/}} \citep{colombo_2015,colombo_2019}. Our approach to the cloud extraction is similar to the one employed in \cite{faustinovieira_2024}. 

The parameters required for the construction of the dendrogram are the minimum threshold for segmentation (\texttt{min\_value}), the minimum difference in emission for two structures to be considered independent (\texttt{min\_delta}), and the minimum area a structure must be (\texttt{min\_npix}). For all galaxies, we adopted $\texttt{min\_value}=5\,\mathrm{M}_\odot \mathrm{pc}^{-2}$ and $\texttt{min\_delta}=5\,\mathrm{M}_\odot \mathrm{pc}^{-2}$ to ensure the extraction is consistent across the sample. We select $\texttt{min\_value}=5\,\,\mathrm{M}_\odot \mathrm{pc}^{-2}$ as this ensures a threshold for detection above the most diffuse emission in the maps. This is done in order to avoid considering noise, but also to help segment each map into more manageable "trunks" to feed to the extraction algorithm, since selecting too low of a threshold will significantly increase computational time. A wide parameter space was explored for \texttt{min\_delta} (as well as \texttt{min\_value}), and we observed minimal difference in the resulting clouds \citep[see also][]{colombo_2015}. The adopted $\texttt{min\_delta}=5\,\mathrm{M}_\odot \mathrm{pc}^{-2}$ value shows consistent segmentation across the whole sample; values much higher than this seemed to result in generally larger clouds in NGC\,628 (given that it has the highest spatial resolution in the sample), whilst the remaining galaxies appeared unaffected. The \texttt{min\_npix} was set to the number of pixels which correspond to 3 resolution elements ($\sim9$~pix per resolution element), to ensure all identified structures are well-resolved. The major difference in the cloud extraction performed here versus that performed in \cite{faustinovieira_2024}, is the use of the "luminosity" (i.e. surface density) as well as the "radius" criterion to aid the \texttt{SCIMES} clustering. Furthermore, we employ a user-defined scaling parameter of 100~pc for the "radius" criterion, which helps the algorithm identify structures in an equal manner across maps of different galaxies which will have different dynamical ranges. This scaling parameter sets an upper threshold on the size of the structures that can be identified by \texttt{SCIMES}, although this is not a hard upper limit.

The \texttt{SCIMES} segmentation resulted in a total of $\sim97,000$ clusters recovered across our galaxy sample (at each galaxy's native resolution), which encompass on average $\sim70\%$ of the total mass in our maps. Since this cloud catalogue is based on extinction-derived maps, both atomic and molecular clouds are included, given that dust traces the total gas. In order to build a fiducial catalogue of clouds for the analysis performed in this paper, a number of further selection criteria were applied to ensure the clouds have robust properties. 

The relative uncertainty of each cloud's surface density is estimated through Monte Carlo simulations (see Appendix~\ref{appA:monte_carlo}). In the fiducial sample, we keep only clouds with a relative uncertainty $<30\%$, which are subsequently flagged in our catalogue with \textit{Rel\_err\_cut}=1. Furthermore, we also determine the maximum measurable limit, $\tau_\mathrm{max}$, at each pixel across all our maps. This quantity is dictated by the photometric noise of the original HST V-band images, and so, remembering Eq.~\ref{eqn:tau_form}, we impose that this maximum is when $I - I_\mathrm{fg} = 3\,\sigma_I$ ($\sigma_I$ being the photometric noise). This is because the maximum value of $\tau$ corresponds to the term $I - I_\mathrm{fg}$ being at a minimum. We determine the fraction of pixels within each cloud where our optical depth estimate surpasses $\tau_\mathrm{max}$, which would reflect higher uncertainty in the surface density/mass estimates for that cloud. The fiducial sample holds only clouds whose fraction of pixels where $\tau>\tau_\mathrm{max}$, relative to the total number of pixels within the cloud, is less than $30\%$. These clouds are flagged in the catalogue with \textit{Tau\_max\_cut}=1. It is important to note that for NGC\,4321 we diverge slightly from this decision, since a considerable amount of clouds in one of the segments of the bar do not pass this \textit{Tau\_max\_cut}. Our assumption of a single background/foreground value for the entire galactic disc is likely failing for this portion of the large bar, since we do not account for the inclination of the bar within a more spherical stellar bulge, which technically means that the near and far sides of the bar are bound to have different foreground to background fractions. Accounting for this in our calibration process is not straightforward and requires more sophisticated modelling of the stellar light in bars, which is beyond the current scope of this work. In an effort to minimise the loss of statistics for bar environments, which is essential to the analysis of impact of environments performed in this paper, we decide to include these bar clouds, but raise an additional flag for the NGC\,4321 catalogue (\textit{Uncertain\_mass\_tag}=1), alerting that the mass estimates for these clouds are more uncertain.  There are 220 of these clouds, constituting $7\%$ of clouds in those environments. All observed trends with cloud properties remain the same with or without these more uncertain clouds in NGC\,4321. Finally, we rule out any clouds that directly touch the image's edge (or inner mosaic edge in the case of NGC\,628).

In this work, we analyse any trends of cloud properties between the different environments and across galactocentric radius within each galaxy, at the "native" physical resolution of the given galaxy (Section~\ref{sec:nat_cat}). We also compare the cloud populations between galaxies, at a common, homogenised resolution (Section~\ref{sec:uni_cat}). For that purpose, we degrade the physical resolution of the extinction-derived maps for NGC\,4689, NGC\,628 and NGC\,4321 (with a Gaussian convolution) to the resolution of NGC\,1566, which is the furthest galaxy in the sample. Therefore, the homogenised gas surface density maps all have a physical resolution of $\sim9$~pc (see Table~\ref{tab:gal_properties}). The cloud extraction process described above is then repeated for these homogenised maps, resulting in a new, homogenised resolution cloud catalogue. All clouds in this homogenised catalogue have sizes larger than $\sim 303$\,pc$^2$, which corresponds to 3 resolution elements at the distance of NGC\,1566, to ensure they are well-resolved. 

In conclusion, in this paper we analyse cloud populations in NGC\,4689, NGC\,628, NGC\,1566 and NGC\,4321 extracted at both the native resolution of each extinction-derived map, as well as at a common physical resolution which is set by the furthest target (NGC\,1566). Clouds in the fiducial sample for either case, native or homogenised resolution, are flagged in the corresponding catalogues with \textit{Rel\_err\_cut}=1, \textit{Tau\_max\_cut}=1, and \textit{Not\_edge\_cut}=1 (see Appendix~\ref{app:catalogue}). The fiducial sample for the native resolution holds $\sim77,000$ clouds across our sample of galaxies (see Fig.~\ref{fig:science_asgn}), whilst the homogenised one is composed of $\sim60,000$ clouds.

\subsection{Cloud properties}

In this work, we analyse any trends between cloud properties and their galactic context. All the derived cloud properties are listed and described in Appendix~\ref{app:catalogue} (see Table~\ref{tab:catalogue}), with only a brief description of a subset of properties relevant for our analysis included below. None of these properties have been de-projected for the galaxy's inclination and position angle.

The average gas mass surface density of each cloud ($\Sigma_\mathrm{avg}$) is estimated from the total "flux" (i.e. total gas mass surface density) computed by \texttt{ASTRODENDRO} for the structure, divided by the footprint area of the cloud. The cloud mass ($M$) is then given by its $\Sigma_\mathrm{avg}$ multiplied by its physical area ($A$ in pc$^2$). Finally, to get an estimate of each cloud's length and elongation, we retrieve the medial axis of each cloud in our catalogue. The medial axis is the longest running spine of a cloud, i.e. the longest continuous line that is the furthest away from the edges of the cloud. Utilising the medial axis is a purely geometrical approach which can be more faithful to the actual size and shape of a resolved cloud \citep[see e.g.][]{duarte-cabral_2021}. In this work, we use the medial axis length ($L_\mathrm{MA}$) as well as the medial axis aspect ratio ($\mathrm{AR}_\mathrm{MA}$), which is $L_\mathrm{MA}$ divided by the width of the cloud (i.e. twice the average distance from the medial axis to the cloud edges).  When this aspect ratio is close to unity, the cloud tends to be circular.

In addition, to categorise clouds in terms of their morphology, we employ the automated technique Rotated $J$-plots, or RJ-plots \citep{jaffa_2018,clarke_2022}. This is a way to quantitatively characterise a cloud's morphology by comparing its principal moments of inertia to those of a circle of equal area and weight\footnote{Weight here is the integral of each pixel's weighting \citep[see][]{clarke_2022}.}. RJ-plots automatically categorise clouds into four classifications: circular (RJ=1), ring-like (RJ=2), elongated and centrally overdense (RJ=3), and elongated and centrally underdense (RJ=4). 

\begin{table}
    \centering
    \begin{tabular}{c c c c c c}

    \hline
    \hline
        
     Env. & $N_\mathrm{clouds}$ &  $M$ & $\Sigma_\mathrm{avg}$  & $L_\mathrm{MA}$ &  AR$_\mathrm{MA}$  \\
    
     &  & ($10^3$~M$_\odot$) & ($\mathrm{M}_\odot \mathrm{pc}^{-2}$) & (pc) & \\
        
    (1) & (2) & (3) & (4) & (5) & (6) \\ 
    
    \hline

    \multicolumn{6}{c}{NGC\,4689} \\

    \hline

    C & 141 & $6.0_{3.4}^{21.2}$ & $11.8_{9.5}^{15.3}$ & $14_{10}^{26}$ & $3.0_{2.25}^{4.24}$ \\
    
    D & 13,212 & $3.2_{2.3}^{5.2}$ & $8.2_{7.6}^{9.0}$ & $13_{10}^{18}$ & $3.0_{2.36}^{4.02}$ \\

    Global & 13,353 & $3.2_{2.3}^{5.3}$ & $8.2_{7.6}^{9.1}$ & $13_{10}^{18}$ & $3.0_{2.36}^{4.02}$ \\

    \hline

    \multicolumn{6}{c}{NGC\,628} \\

    \hline

    C & 393  & $3.1_{1.6}^{9.6}$ & $8.7_{7.6}^{11.1}$ & $17_{12}^{37}$ & $3.29_{2.36}^{5.25}$ \\
    
    SA & 9,733 & $1.7_{1.1}^{3.2}$ & $8.2_{7.6}^{8.9}$ & $14_{10}^{21}$& $3.13_{2.36}^{4.38}$ \\
    
    IA & 16,068 & $1.6_{1.1}^{3.1}$ & $8.2_{7.7}^{8.8}$ & $14_{11}^{22}$ & $3.2_{2.4}^{4.5}$ \\

    Global & 26,194 & $1.7_{1.1}^{3.2}$ & $8.2_{7.6}^{8.8}$ & $14_{11}^{22}$ & $3.16_{2.37}^{4.47}$ \\
    
    \hline 

    \multicolumn{6}{c}{NGC\,1566} \\

    \hline

    C/B & 627 & $10.9_{5.4}^{34.5}$ & $10.3_{8.15}^{16.0}$ & $16_{10.5}^{29}$ &  $3.13_{2.25}^{4.53}$ \\
    
    SA & 4,268 & $7.5_{4.2}^{17.6}$ & $9.4_{8.1}^{13.4}$ & $16_{10.5}^{29}$ & $3.0_{2.25}^{4.25}$ \\
    
    IA & 15,711 & $5.7_{3.6}^{11.5}$ & $8.6_{8.0}^{9.5}$ & $14_{10}^{22}$ & $3.2_{2.37}^{4.59}$ \\

    Global & 20,606 & $6.1_{3.7}^{12.9}$ & $8.7_{8.0}^{9.9}$ & $14_{10}^{22}$ & $3.13_{2.36}^{4.5}$ \\
    
    \hline

    \multicolumn{6}{c}{NGC\,4321} \\

    \hline

    C/B & 2,446 & $7.1_{3.6}^{16.6}$ & $12.2_{9.4}^{19.3}$ & $14_{10}^{22}$ & $2.93_{2.25}^{4.02}$ \\
    
    SA & 5,314 & $4.6_{2.9}^{9.3}$ & $9.9_{8.8}^{12.2}$ & $13_{10}^{19}$ & $3.0_{2.36}^{4.0}$ \\
    
    IA & 8,885 & $4.2_{2.7}^{8.1}$ & $9.3_{8.5}^{10.6}$ & $13_{10}^{20}$ & $3.0_{2.36}^{4.25}$ \\

    Global & 16,645 & $4.6_{2.9}^{9.5}$ & $9.7_{8.6}^{11.7}$  & $13_{10}^{20}$ & $3.0_{2.36}^{4.1}$ \\
    
    \hline
         
    \end{tabular}
    \caption{Characteristics of the cloud populations of each galaxy at their respective physical resolution (native sample). (1) Galactic environment: C(/B)=Centre(/Bar), SA=Spiral arms, IA=Inter-arm, or D=Disc in the case of NGC\,4689. (2) Number of clouds per environment, $N_\mathrm{clouds}$. (3) Cloud mass, $M$. (4) Average gas surface density of clouds, $\Sigma_\mathrm{avg}$. (5) Medial axis length, $L_\mathrm{MA}$. (6) Moment aspect ratio, $\mathrm{AR}_\mathrm{MA}$. For columns (3)-(6), the median of the distribution is presented, with the 25$^\mathrm{th}$ and 75$^\mathrm{th}$ percentiles being the subscript and superscript, respectively.}
    \label{tab:native_sample}
\end{table}

\section{Cloud trends within host galaxy}
\label{sec:nat_cat}

\begin{figure*}
    \centering
    \includegraphics[width=\textwidth]{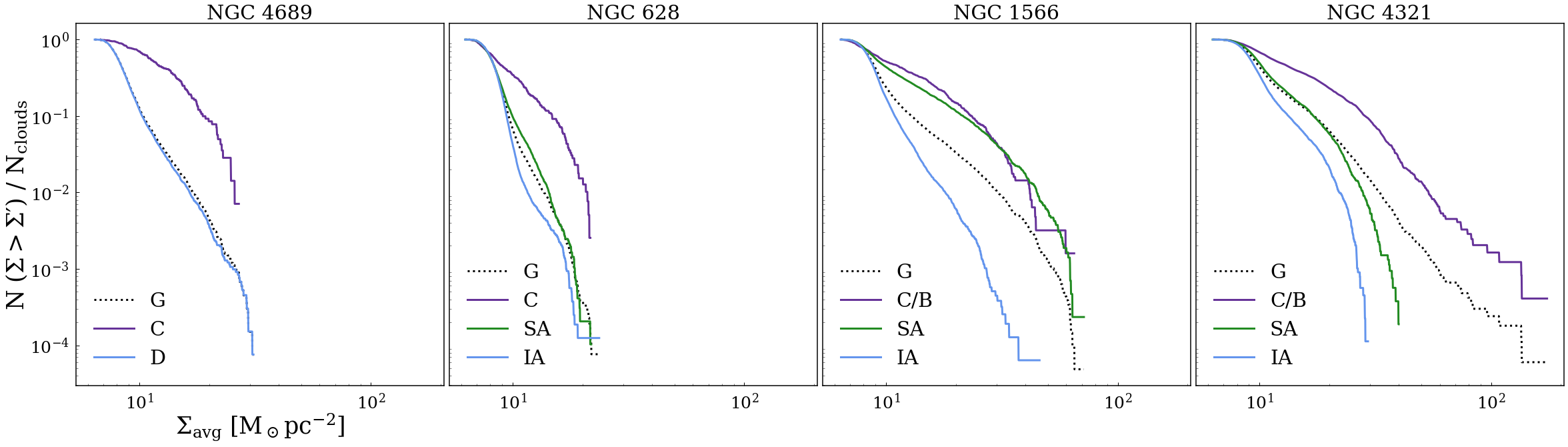}
    \vspace{-0.5cm}
    \caption{Cumulative average cloud surface density distributions across the different environments within our sample of galaxies: NGC\,4689, NGC\,628, NGC\,1566 and NGC\,4321 (from left to right). The different environments are colour-coded, with the centre/bar (C/B) in purple, the spiral arms (SA) in green, the inter-arm regions (IA) in blue, and the global (G) distribution being the dotted black line. For NGC\,4689, the blue denotes the disc (D). All distributions are normalised by the total number of clouds within each relevant environment, $N_\mathrm{clouds}$ (as listed in Table~\ref{tab:native_sample}).}
    \label{fig:CMD_native}
\end{figure*}

\begin{figure}
    \centering
    \includegraphics[width=0.45\textwidth]{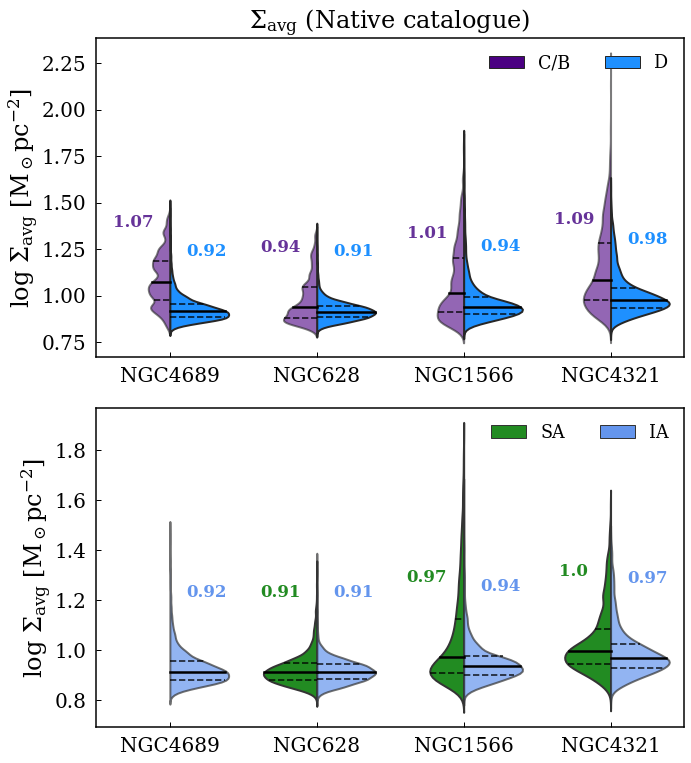}
    \caption{Violin plots showing the contrast between the average cloud surface density ($\Sigma_\mathrm{avg}$) distributions in different environments across the 4 galaxies (native resolution). For each property, the top row shows the difference between the centre/bar (C/B, in purple) and the disc (D, in blue) populations, whilst the bottom row shows the arm population (SA, in green) against the inter-arm (IA, in light blue). For all distributions, the solid black line within the violin represents the median, whilst the dashed lines depict the interquartile range. Next to each distribution, the written label is the relevant median (in logarithmic scale), colour-coded by environment.}
    \label{fig:splitvp_native_sd}
\end{figure}

In this Section, we take each cloud catalogue obtained at the relevant galaxy's physical resolution, and we explore any existing trends of cloud properties as a function of both galactic environment and galactocentric distance. The galactocentric distances used here have been projected onto the plane of the sky, taking into account each galaxy's inclination and position angle (listed in Table~\ref{tab:gal_properties}).

\begin{figure*}
    \centering
    \includegraphics[width=\textwidth]{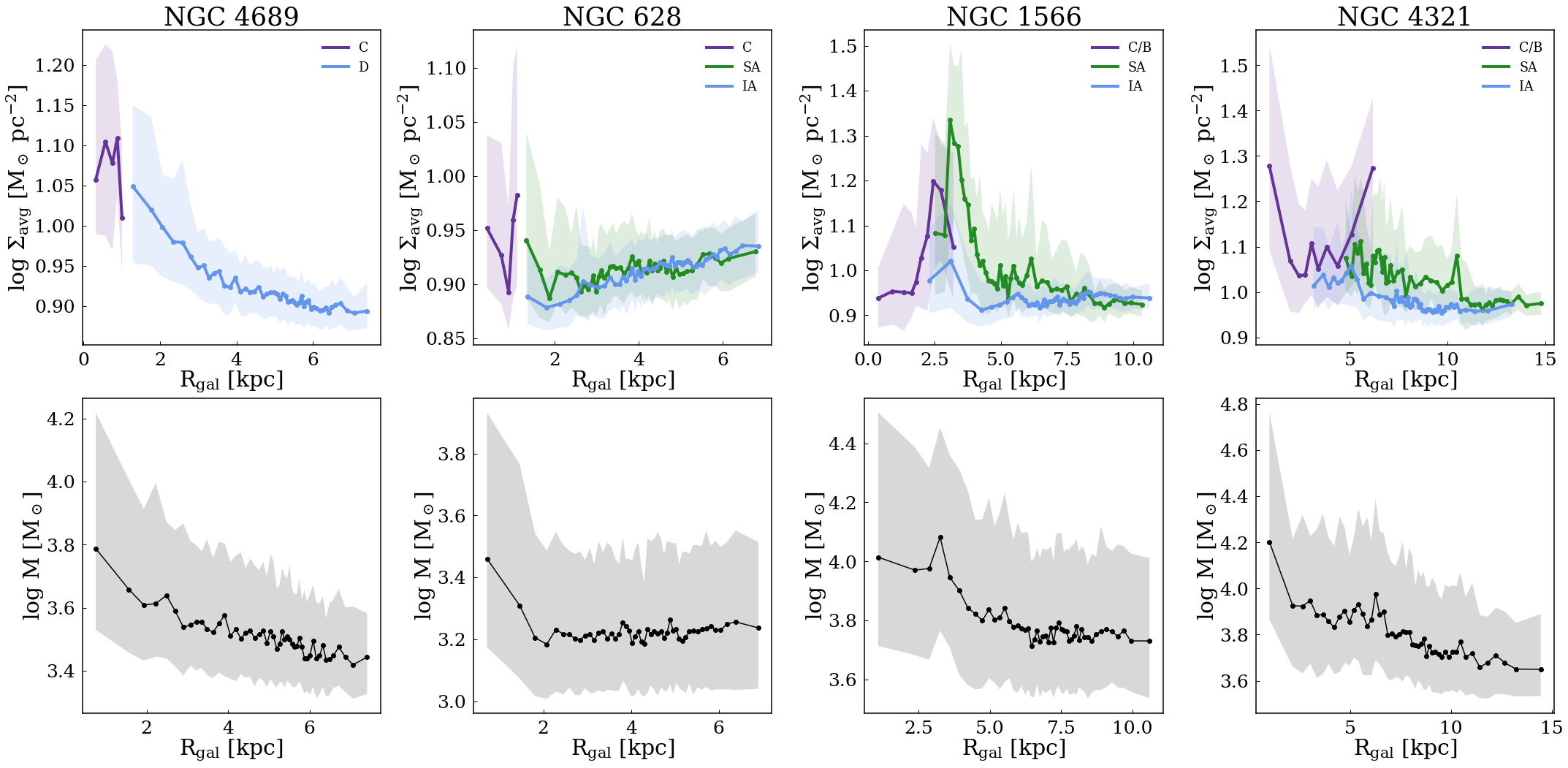}
    \caption{\textit{Top row}: Average cloud surface density ($\Sigma_\mathrm{avg}$) as a function of galactocentric distance for (from left to right) NGC\,4689, NGC\,628, NGC\,1566 and NGC\,4321. The different galactic environments are colour-coded as purple for the centre/bar, green for the spiral arms, or blue for the inter-arm regions (disc in the case of NGC\,4689). \textit{Bottom row:} Cloud mass ($M$) as a function of galactocentric distance. For all panels, the solid line depicts the running median, whilst the shaded regions represent the interquartile range of the relevant distribution. All bins within each respective environment are equally populated.}
    \label{fig:radial_native_sd}
\end{figure*}

\begin{figure}
    \centering
    \includegraphics[width=0.47\textwidth]{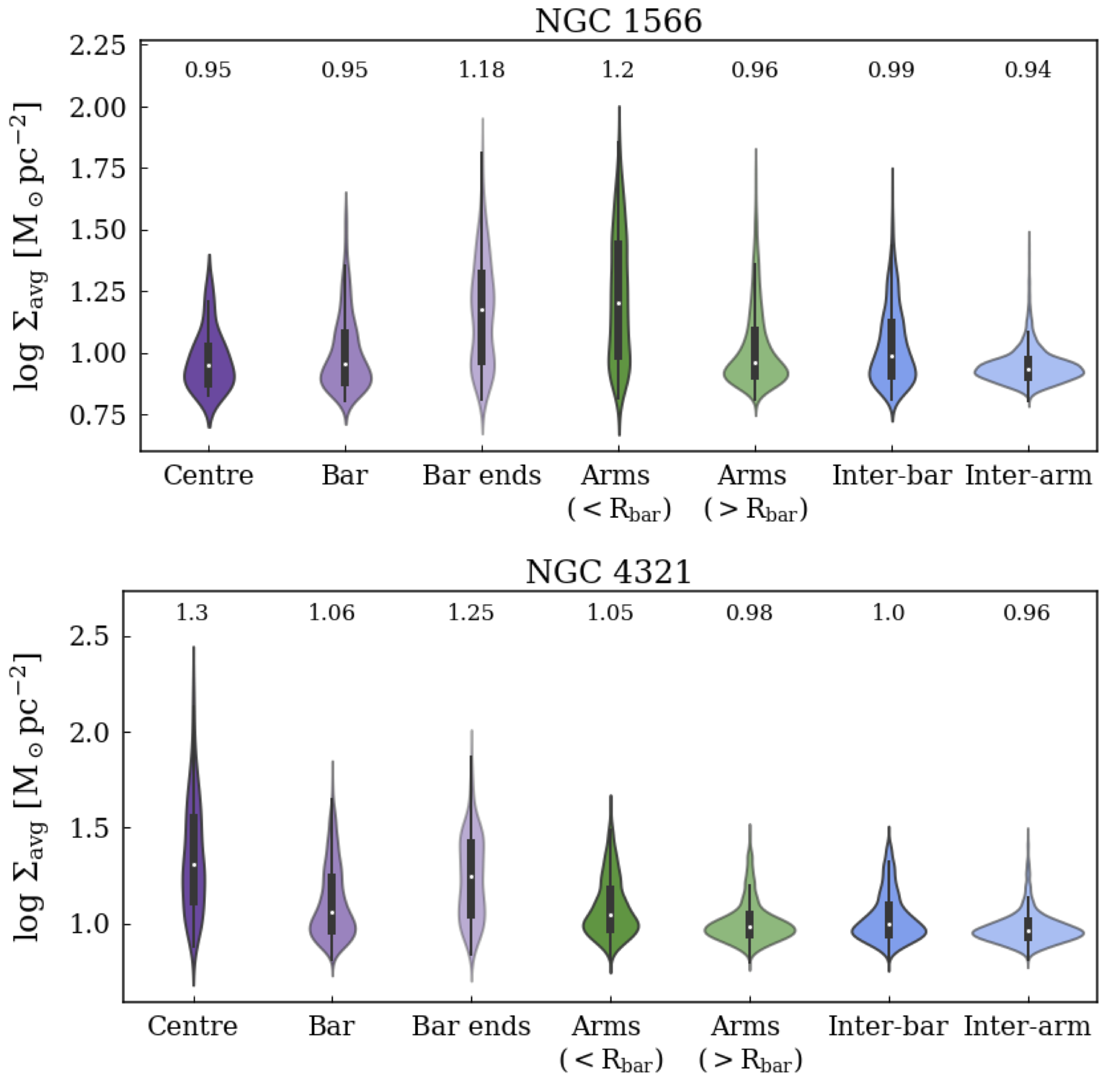}
    \caption{Violin plots showing the average cloud surface density distributions for the barred galaxies in the sample: NGC\,1566 (\textit{top}) and NGC\,4321 (\textit{bottom}). The centre/bar, spiral arms and inter-arm total environments were subdivided into centre, bar, bar ends, arms within the bar radius ($\mathrm{R}_\mathrm{bar}$), arms outside of $\mathrm{R}_\mathrm{bar}$, inter-bar (inter-arm within $\mathrm{R}_\mathrm{bar}$) and inter-arm, as defined in \citet{querejeta_2021}. For each violin plot, the black box depicts the interquartile range of the relevant distribution, with the white dot representing the median (which is also written in logarithmic form above each violin plot).}
    \label{fig:vp_barred_native}
\end{figure}

\subsection{Cloud masses and surface densities}
\label{sec:nat_sd}

From the cumulative $\Sigma_\mathrm{avg}$ distributions shown in Fig.~\ref{fig:CMD_native}, it is evident that the central cloud surface densities are distributed differently than disc clouds within each respective galaxy. This is further exemplified in Fig.~\ref{fig:splitvp_native_sd}, which also shows that cloud surface densities are consistently higher in the centres/bars of galaxies versus the disc - a finding consistent with the literature \citep[e.g.][although this could also be partially driven by cloud superposition along the line-of-sight]{sun_2020, querejeta_2021, faustinovieira_2024}. This behaviour is also seen with cloud masses. Figures~\ref{fig:CMD_native} and \ref{fig:splitvp_native_sd} also showcase some difference between cloud populations in the arm versus inter-arm environments across the sample of spiral galaxies, which is much more pronounced in the two barred targets.

Figure~\ref{fig:radial_native_sd} illustrates the running radial medians of $\Sigma_\mathrm{avg}$ (within each environment) and cloud mass, $M$, for the present sample of galaxies. For the cloud mass, we show only the radial profile of all clouds in each galaxy, as the trends between environments are similar to those observed with average surface density. To construct the total radial profiles, we divide each galaxy into 50 equally populated concentric radial bins. To look at radial trends within each large-scale environment, we instead create 50 bins for the disc environments, and either 5 (in the case of NGC\,4689 and NGC\,628) or 10 radial bins (for NGC\,1566 and NGC\,4321) for the centre/bar environments, in an attempt to hold at least 50 clouds in each bin. The radial bins within each large-scale environment are also equally populated to minimise binning biases. The observed trends in each galaxy are discussed in more detail below.

For the barred galaxies of the sample (NGC\,1566 and NGC\,4321) we further divide the cloud populations into the "sub-environments" defined in \cite{querejeta_2021} (see Fig.~\ref{fig:vp_barred_native}). These include the centre, bar, and bar ends (defined as the overlap between the bar and start of the spiral arms), as well as the arm and inter-arm environments inside and outside the bar radius ($\mathrm{R}_\mathrm{bar}$).

\subsubsection{NGC\,4689}

In Fig.~\ref{fig:CMD_native}, it is possible to see that although the centre and disc cumulative distributions of NGC\,4689 extend to roughly the same $\Sigma_\mathrm{avg}$ values, the amount of high surface density clouds is much higher in the centre than in the disc. This ties into the behaviour shown in Fig.~\ref{fig:splitvp_native_sd}, where clouds in the disc show a more bottom-heavy distribution (in terms of surface density), relative to the centre. In other words, there are more lower surface density clouds in the disc than in the centre. This is also seen, albeit less pronounced, with cloud masses. For this flocculent galaxy, average cloud surface densities (and masses) generally decrease with larger galactocentric distance, as can be seen from Fig.~\ref{fig:radial_native_sd}.

\subsubsection{NGC\,628}

NGC\,628 also displays a higher concentration of high-$\Sigma_\mathrm{avg}$ clouds in the centre versus the disc (Fig.~\ref{fig:splitvp_native_sd}), although the difference is less obvious as that seen in NGC\,4689. From Figs.~\ref{fig:CMD_native} and \ref{fig:splitvp_native_sd}, we can also see that there are very little differences between the inter-arm and spiral arm populations in NGC\,628.

Figure~\ref{fig:radial_native_sd} shows that in NGC\,628 there is a peak of average cloud surface densities at the radii where the centre meets the spiral arms. It also appears that for small galactocentric radii ($\mathrm{R}_\mathrm{gal}\lesssim2.5$~kpc), the clouds in the spiral arms show relatively higher values of $\Sigma_\mathrm{avg}$ compared to inter-arm clouds (particularly evident when looking at the interquartile spread). Past this radius, the disc seems to behave in similar fashion, with minimal differences between the two environments, translating in the very similar distributions seen in Fig.~\ref{fig:splitvp_native_sd}. It is important to note that although in Fig.~\ref{fig:radial_native_sd} there seems to be a slight rise in $\Sigma_\mathrm{avg}$ towards larger $\mathrm{R}_\mathrm{gal}$, this trend is not likely to be significant, as the increase is of the order $\sim0.04$\,dex (within the observed interquartile scatter), and the retrieval of clouds becomes more incomplete towards larger radii. The cloud mass radial profile shows a decrease up until $\mathrm{R}_\mathrm{gal}\sim2$~kpc, after which it seems to flatten.

\subsubsection{NGC\,1566}

NGC\,1566 is one of the barred spiral galaxies in the sample, with a bar that extends for $\sim6$~kpc in diameter, and very bright, well-defined arms, as can be seen from Fig.~\ref{fig:sd_sample}. Unlike the other galaxies studied here, the $\Sigma_\mathrm{avg}$ cumulative distribution for the arms is remarkably similar to that of the centre/bar environment. The spiral arms in this galaxy also display a considerable offset from the distribution of the inter-arm clouds. This is also seen in Fig.~\ref{fig:splitvp_native_sd}, where the arm population appears more top-heavy in comparison to the inter-arm, with a very extended tail towards higher values. In the present sample of galaxies, NGC\,1566 displays the largest arm/inter-arm difference in average cloud surface densities.

In terms of radial trends, Fig.~\ref{fig:radial_native_sd} shows that NGC\,1566 has a peak of surface density at $\mathrm{R}_\mathrm{gal}=2$ -- $4$~kpc (i.e. towards the end of the bar and start of spiral arms). There is an increase of $\Sigma_\mathrm{avg}$ as $\mathrm{R}_\mathrm{gal}$ increases within the central environments of NGC\,1566. Both the spiral arms and inter-arm clouds show high values of $\Sigma_\mathrm{avg}$ towards the end of the bar at $\mathrm{R}_\mathrm{gal}\sim3$~kpc (although this is much more pronounced for the arms), after which average cloud surface densities decrease. Furthermore, arm clouds consistently present higher surface densities up until $\sim8$~kpc, after which the arm and inter-arm distributions become similar. These radial behaviours are consistent with the differences seen between these two environments in Fig.~\ref{fig:splitvp_native_sd} (and Fig.~\ref{fig:CMD_native}). In terms of cloud mass, there seems to be a steady decline with $\mathrm{R}_\mathrm{gal}$ in the global radial profile, with a sudden spike around the transition between centre/bar and disc ($\mathrm{R}_\mathrm{gal}=2$ -- $4$~kpc). This is expected, since cloud-cloud collisions seem to be enhanced towards bar ends \citep[e.g.][]{fujimoto_2020,maeda_2023, maeda_2025}.

Figure~\ref{fig:vp_barred_native} illustrates the $\Sigma_\mathrm{avg}$ distributions for the different sub-environments within NGC\,1566. Clouds appear to have similar distributions of $\Sigma_\mathrm{avg}$ across the centre, bar, and inter-bar regions, with a notable change toward the bar ends and the inner-most parts of the spiral arms, where the distributions are much more spread out, reaching much higher values. This enhances the idea that the bar ends, at $\sim3$~kpc, are promoting a very different environment (which is conducive to cloud growth) compared to the rest of the disc in NGC\,1566, in line with what we had seen from the observed radial trends in this galaxy (Fig.~\ref{fig:radial_native_sd}). Outside $\mathrm{R}_\mathrm{bar}$, the SA and IA distributions become more similar and appear more bottom-heavy for this galaxy.

\subsubsection{NGC\,4321}

As can be seen from Fig.~\ref{fig:CMD_native}, the distribution of cloud surface densities in the centre/bar of NGC\,4321 (which has a large bar, extending for $\sim8.7$~kpc in diameter) extends past the arms and inter-arm into much higher surface densities. This translates into the centre/bar displaying higher surface densities than the disc, in line with what is seen for the rest of the sample. In Fig.~\ref{fig:splitvp_native_sd} we can also see some differences in the distribution of the arm and inter-arm cloud populations, with clouds in the arms showcasing slightly higher values of $\Sigma_\mathrm{avg}$. 

From Fig.~\ref{fig:radial_native_sd}, we can see that NGC\,4321 shows increased cloud surface densities towards the centre and at the very end of the bar, with little variation in the remainder of the galactic disc. Similar to NGC\,1566, clouds seem to have higher values of $\Sigma_\mathrm{avg}$ in the spiral arms than their inter-arm counterparts, up until a certain radius ($\sim11$~kpc), after which they become similar. Overall, this galaxy displays a steady decline of cloud masses with increasing galactocentric radius.

Strikingly, the centre/bar radial profiles of NGC\,1566 and NGC\,4321 are very different, which could suggest different dynamics and bar-driven gas flows. However, further study of radial gas flows in these galaxies (from either an observational or numerical perspective) is needed for any definitive conclusions to be drawn \citep[e.g.][]{wong_2004,querejeta_2016}. Still, as can be seen from Fig.~\ref{fig:vp_barred_native}, clouds do not display the same behaviour within the centre, bar and bar ends environments as what is seen in NGC\,1566. The centre environment harbours the highest surface density clouds, whilst clouds along the bar display a more bottom-heavy distribution of surface densities\footnote{This behaviour remains unchanged if the clouds in NGC\,4321 with more uncertain masses are not considered, as explained in Section~\ref{sec:cloud_extraction}.}. There is still a rise in $\Sigma_\mathrm{avg}$ towards the bar ends, although less pronounced than in NGC\,1566. Similarly, we see that the inner spiral arms and inter-bar environments are also displaying higher $\Sigma_\mathrm{avg}$ in NGC\,4321, whilst the environments at $>\mathrm{R}_\mathrm{bar}$ appear more bottom-heavy, although the differences are more subtle than in NGC\,1566. This suggests that the dynamics driven by bars are affecting the environments within their radius.

\begin{figure*}
    \centering
    \includegraphics[width=\textwidth]{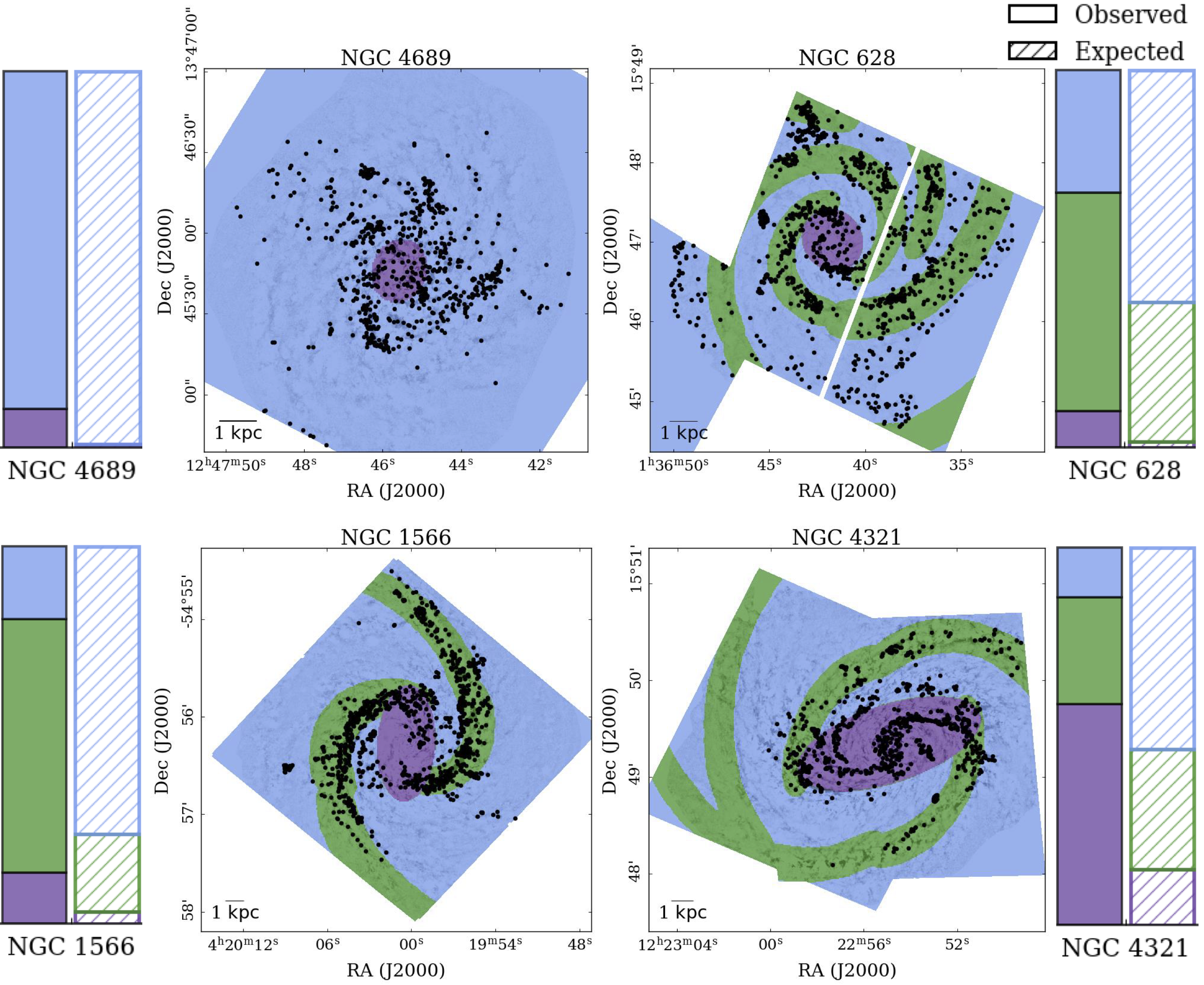}
    \caption{Spatial distribution of the top 5\% clouds in terms of average surface density ($\Sigma_\mathrm{avg}$) across the sample, represented as black dots. For each galaxy, the background greyscale is the gas surface density map, and the environmental masks are in colour (purple for centre/bar, green for spiral arms, and blue for inter-arm or disc in NGC\,4689). Next to each subplot are the respective stacked bar charts for that galaxy, showing the observed environmental distribution of these extreme clouds (solid) versus the expected number of clouds per environment (hatched, see Table~\ref{tab:native_sample}).}
    \label{fig:extreme_sd_native}
\end{figure*}

\subsubsection{Extreme clouds: high $\Sigma_\mathrm{avg}$}
\label{sec:extreme_sd}

It is also possible to isolate the highest surface density objects of each galaxy and investigate their preferred location within the large-scale galactic context. If there is a higher than expected concentration of high-$\Sigma_\mathrm{avg}$ clouds towards a given environment, it could hint at some physical process that benefits the formation and growth of clouds in that region. One such process could be an increased frequency of cloud-cloud collisions \citep[e.g.][]{dobbs_2008,inutsuka_2015}. Figure~\ref{fig:extreme_sd_native} showcases the positions and environmental distributions of these high-surface density clouds within the respective galaxy. For each galaxy, these clouds compose the top 5\% of the $\Sigma_\mathrm{avg}$ distribution. This subsample constitutes 667 clouds in NGC\,4689, $1309$ in NGC\,628, $1030$ in NGC\,1566 and 832 in NGC\,4321. It could be that the distribution of these high-$\Sigma_\mathrm{avg}$ across the large-scale environments is not significant, and is just an effect of random sampling. To test that, a comparison is performed between the fraction of these clouds per environment and the fraction that is expected from the general all-galaxy distribution. This is done through a Pearson $\chi^2$ statistical analysis:
\begin{equation}
    \chi^2 = \sum^{n}_{i=1} \frac{(O_i - E_i)^2}{E_i},
\end{equation}
where $n$ is the number of categories (environments, in this case), $O_i$ is the number of observed counts, and $E_i$ is the number of expected counts from the theoretical distribution (i.e. the all-galaxy distribution, see Table~\ref{tab:native_sample}). We perform $10^6$ random draws (without replacement) of a subset of $N$ clouds, calculating a $\chi^2$ value for each draw. This allows us to build a cumulative distribution across the $10^6$ iterations, thus giving an estimated likelihood ($\mathrm{p}_\mathrm{rnd}$) of a given $\chi^2$ value. We then compare the observed $\chi^2$ of the extremely high-$\Sigma_\mathrm{avg}$ subsample of clouds against this simulated distribution. If the $\mathrm{p}_\mathrm{rnd}$ associated to the observed $\chi^2$ value is low, then the relevant distribution is unlikely to arise from random sampling, and thus the observed behaviour is likely to be significant rather than a by-product of sampling. In other words, the large-scale environment may have a role in promoting these specific types of clouds. There is no attempt to perform this categorical $\chi^2$ test on NGC\,4689, since for this galaxy there are only 2 categories (i.e. environments) with a huge discrepancy in population. This means that when performing random draws, it is highly likely to pull only clouds from the disc, which makes determining a $\chi^2$ value not realistic.

\begin{figure}
    \centering
    \includegraphics[width=0.45\textwidth]{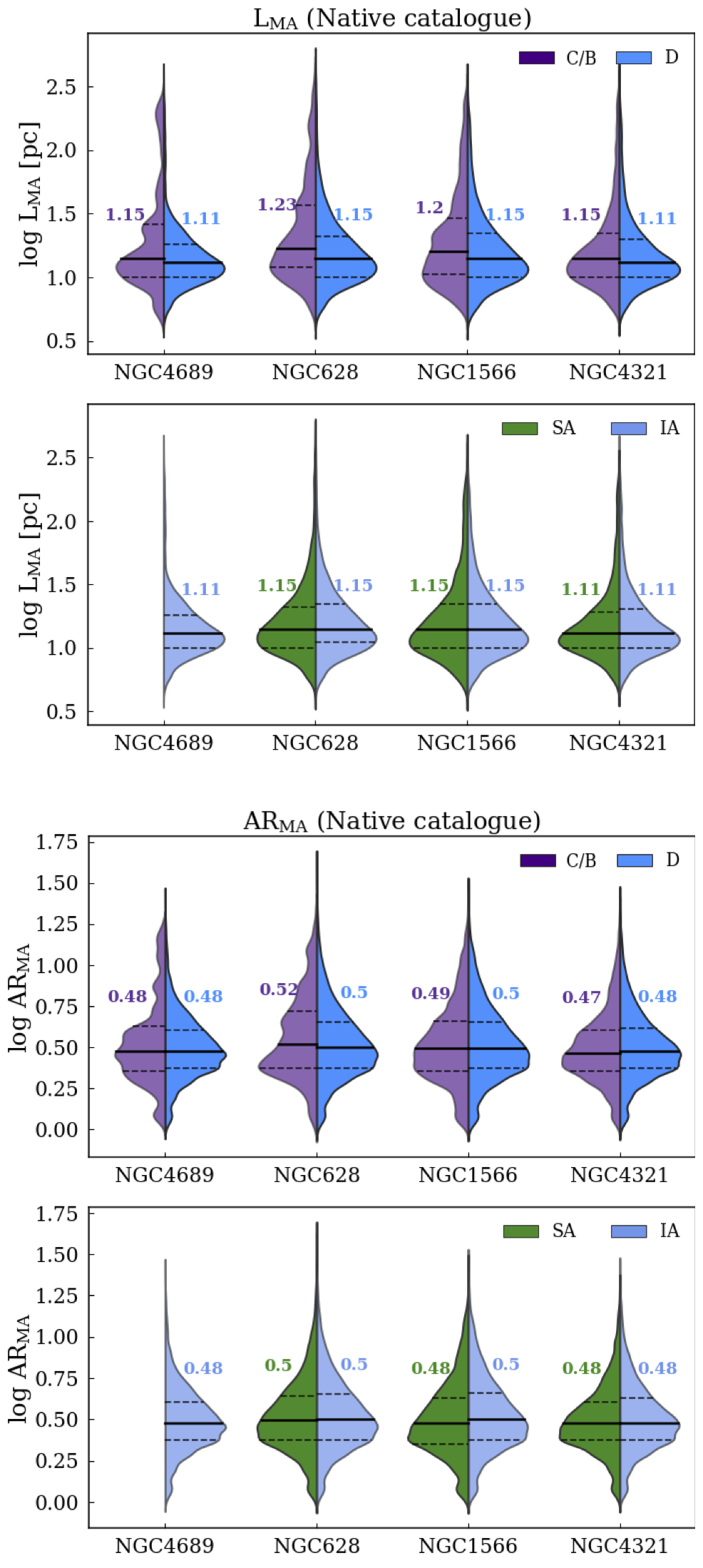}
    \caption{Same as Fig.~\ref{fig:splitvp_native_sd}, for cloud size ($L_\mathrm{MA}$, \textit{top}) and aspect ratio ($\mathrm{AR}_\mathrm{MA}$, \textit{bottom}).}
    \label{fig:splitvp_native_L_AR}
\end{figure}

Visually, it appears that the flocculent NGC\,4689 harbours most of its high-$\Sigma_\mathrm{avg}$ clouds at small galactocentric radii, which is consistent with the radial trends (see Fig.~\ref{fig:radial_native_sd}). As for the remaining galaxies, Fig.~\ref{fig:extreme_sd_native} shows these extreme clouds tracing the arms, with also some concentration towards the centres/bars. NGC\,1566 displays many high-$\Sigma_\mathrm{avg}$ objects along the arms, and very little in the inter-arm. This behaviour is consistent with the stark arm/inter-arm contrast seen in Fig.~\ref{fig:splitvp_native_sd}. There are also several extremely high surface density clouds towards the ends of the bar and beginning of spiral arms of NGC\,1566, which is also shown as a peak in the radial $\Sigma_\mathrm{avg}$ profile (Fig.~\ref{fig:radial_native_sd}). As is clear from Fig.~\ref{fig:extreme_sd_native}, in  NGC\,4321, the bar is heavily populated by high surface density clouds. In fact, most of these extreme clouds are located within the central/bar region of the galaxy, with the few objects in the disc being mostly located in the inner galaxy ($<\mathrm{R}_\mathrm{bar}$). This behaviour is unlike that of the other spirals, which show many high-$\Sigma_\mathrm{avg}$ objects along the spiral arms.

For all galaxies considered, high values of $\chi^2$ were obtained, all with $\mathrm{p}_\mathrm{rnd}<10^{-5}$ (see Table~\ref{tab:chi2_sd_lsf}). This is true both when considering the "total" large-scale environments (i.e. C, SA, IA) and when performing the test across the sub-environments in the case of the barred galaxies (as in Fig.~\ref{fig:vp_barred_native}, per \citealt{querejeta_2021}). This test confirms that these extreme clouds with high-$\Sigma_\mathrm{avg}$ are not likely to be randomly distributed. This suggests that some large-scale environments offer physical conditions which are more conducive to the formation of high surface density objects than others. Additionally, the fact that these high-density clouds are not located in the same environments within each of the galaxies suggests that the specific galactic dynamics may also be playing a determining role in the formation of such objects. However, it could be that these high-$\Sigma_\mathrm{avg}$ objects are not individual clouds but rather the superposition of overlapping clouds along the line-of-sight. With the present dataset, it is not possible to determine the contribution of these superposition effects on the observed trends of cloud mass/surface density, as that would require velocity information (with resolution comparable to the scales probed here) to separate overlapping structures. Still, for our largest clouds, it would be possible to check how severe this superposition effect might be with high-resolution ALMA data \citep[e.g with PHANGS-ALMA][]{leroy_2021}, although this is beyond the scope of this work.

\subsection{Cloud sizes and morphology}
\label{sec:nat_morphology}

In this Section, we analyse any trends present in cloud sizes and morphology (either from aspect ratio or RJ-class) within the native resolution cloud catalogue.  

\begin{figure}
    \centering
    \includegraphics[width=0.45\textwidth]{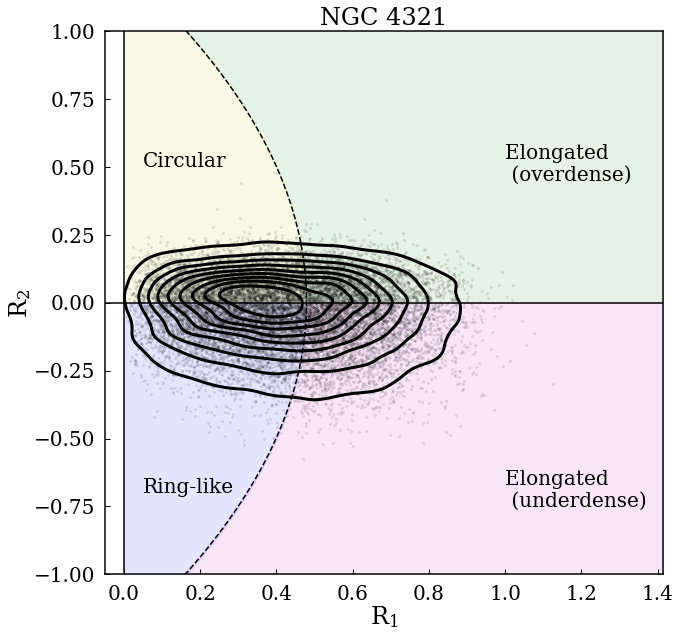}
    \caption{RJ-plot for all clouds in NGC\,4321 (native resolution). The black contours represent the 2D kernel density estimation (KDE) of the underlying distribution of clouds.}
    \label{fig:RJ_NGC4321}
\end{figure}

\begin{figure*}
    \centering
    \includegraphics[width=\textwidth]{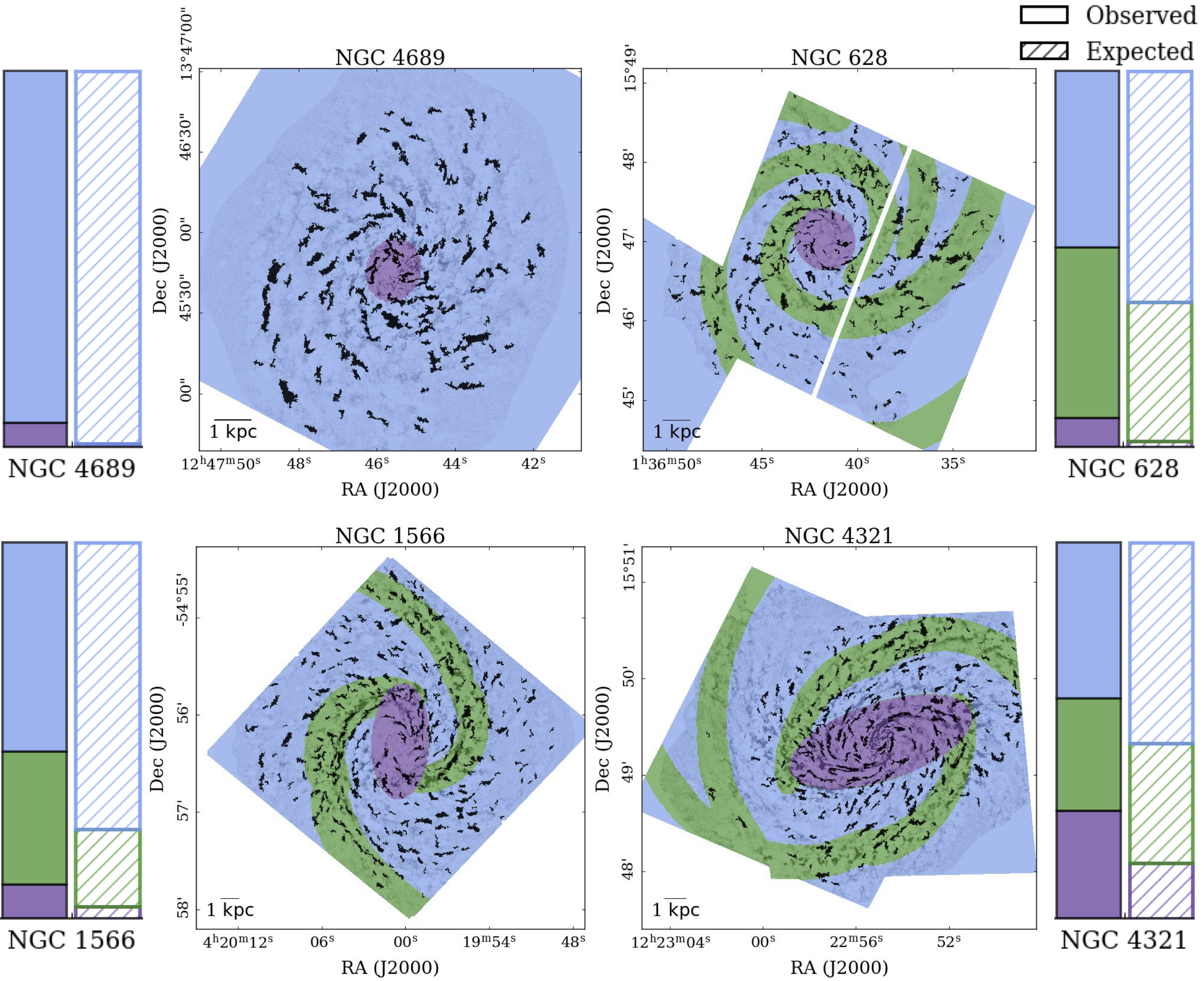}
    \caption{Spatial distribution of "large-scale filaments" (LSFs), which are elongated (RJ=3 or 4) and simultaneously very long ($L_\mathrm{MA}>100~\mathrm{pc}$) clouds. For each galaxy, the background greyscale is the gas surface density map, and the environmental masks are in colour (purple for centre/bar, green for spiral arms, and blue for inter-arm or disc in NGC\,4689). Next to each subplot are the respective stacked bar charts for that galaxy, showing the observed environmental distribution of these extreme clouds (solid) versus the expected number of clouds per environment (hatched, see Table~\ref{tab:native_sample}).}
    \label{fig:LSF_native}
\end{figure*}

\subsubsection{Cloud length and aspect ratio}
\label{sec:lma_arma_nat}

Figure~\ref{fig:splitvp_native_L_AR} displays the medial axis length ($L_\mathrm{MA}$) and aspect ratio ($\mathrm{AR}_\mathrm{MA}$) of clouds across the sample (and within large-scale environments). As is evident from the figure, there is no obvious trend of cloud size between environments across the sample (although there is some slight skewness towards higher values in the centres of NGC\,4689 and NGC\,628). Similarly, there is no discernible trend of aspect ratio across the cloud populations, with all galaxies presenting generally elongated clouds ($\mathrm{AR}_\mathrm{MA}\sim3$, see Table~\ref{tab:native_sample}). As is already hinted from the similarity of the centre/bar and disc populations in Fig.~\ref{fig:splitvp_native_L_AR}, there are no observed radial trends of cloud size or aspect ratio across the sample of galaxies.

\subsubsection{Cloud morphology}
\label{sec:rj_nat}

It is important to note however, that an aspect ratio metric is not the most conclusive. Although the medial axis approach does a better job at identifying truly elongated clouds than the moment-based aspect ratio \citep[see][]{neralwar_2022,faustinovieira_2024}, it will still confuse "straight", filamentary clouds with ring-like clouds, as they can have similar aspect ratios despite being vastly different morphologies. Filamentary clouds could be a sign of stretching due to shear \citep[e.g.][]{koda_2009}, whilst ring-like "bubbles" are usually associated with stellar feedback \citep[e.g.][]{barnes_2023}\footnote{It is important to note, however, that these bubbles are not likely to be formed of one single cloud, but likely many arc-shaped objects.}. Given that these shapes of clouds are likely driven by different physical processes, it is important to disentangle them. 

Figure~\ref{fig:RJ_NGC4321} displays an example of an RJ-plot for NGC\,4321, showcasing that clouds in this galaxy tend to be quasi-circular (i.e. either circular or ring-like). Indeed, in NGC\,4321 roughly 33\% of clouds appear ring-like, and 28\% are circular clouds, whilst elongated clouds make up 39\% of the population. The remaining galaxies in the sample display similar trends. The RJ-plots for the other galaxies (as well as for the different environments within each galaxy) are presented in Appendix~\ref{app:catalogue}. There is no apparent trend of RJ distribution across the environments within each galaxy, as all the distributions look similar (Fig.~\ref{fig:RJ_gal}, see also Table~\ref{tab:RJ_class}).

\subsubsection{Extreme clouds: LSFs}
\label{sec:extreme_lsf}

We also isolate clouds that could approximate large-scale filaments (LSFs) in the sample, shown in Fig.~\ref{fig:LSF_native}. These are clouds belonging to the elongated RJ classes (RJ=3 or 4), with $L_\mathrm{MA}>100~\mathrm{pc}$. There are 126 of such objects for NGC\,4689, 265 for NGC\,628, 309 for NGC\,1566, and 272 for NGC\,4321. From Fig.~\ref{fig:LSF_native}, it appears that the orientation of these LSFs is more "ordered" in NGC\,1566 and NGC\,4321, whilst in NGC\,4689 and especially NGC\,628 their orientation seems more random. In other words, the orientation (and orderliness) of these filaments appears to be related to the underlying morphology, which could be linked to the strength of the spiral potential. In fact, \cite{querejeta_2024} measured the arm/inter-arm contrast of stellar mass surface density for PHANGS galaxies, and found much larger arm/inter-arm contrasts in NGC\,1566 and NGC\,4321 than in NGC\,628, which implies that NGC\,628's spiral potential is much weaker relative to the barred galaxies. In the barred targets, the LSFs in the inter-arm (i.e. akin to inter-arm "spurs"; e.g. \citealt{la-vigne_2006}) appear to become more aligned with the arm as they approach it \citep[i.e. on the concave side, see also][]{duarte-cabral_2016,duarte-cabral_2017}. On the convex side of the arms (i.e. downstream), where the gas would be exiting the arm, these LSFs appear nearly perpendicular to the arm. This behaviour is particularly visible in NGC\,1566. This could be a sign of a stronger influence of the spiral dynamics exerted on the ISM of these galaxies, given that this behaviour is not seen in NGC\,628, which has visually fainter arms. Still, for more firm conclusions, these visual trends need to be confirmed by studying the relative orientation of these filaments with respect to the nearest arm in a more quantitative way. 


As was done with the extremely high surface density clouds in the previous Section, the fraction of observed LSFs for each environment is compared to what is expected from the general all-galaxy distribution, to investigate if the observed distributions show any significant environmental trends. This is done through a Pearson $\chi^2$ statistical analysis\footnote{Again, given the reduced number of environments in NGC\,4689, this analysis is not performed for this galaxy (see Section~\ref{sec:extreme_sd}).}. Large values of $\chi^2$ were obtained across the sample for the galactic centres/bars, all with very low likelihoods ($\mathrm{p}_\mathrm{rnd}<10^{-6}$) of being a by-product of random sampling (see Table~\ref{tab:chi2_sd_lsf}). This suggests that all spiral galaxies in the sample seem to show a higher concentration of LSFs in their centres/bars than expected (see Fig.~\ref{fig:LSF_native}). For all galaxies, there does not seem to be any particularly strong enhancement of LFSs towards spiral arms (with the exception of perhaps NGC\,1566) but there is a systematic trend of a slightly lower amount of observed LFSs in the inter-arm regions than expected (Fig.~\ref{fig:LSF_native}), albeit not very statistically significant. As mentioned above, further study is required to determine the physical underpinnings of the observed trends, as these are likely associated with galactic dynamics, and in particular the combination of shear and distortion from the galactic potential.

\begin{table}
    \centering
    \begin{tabular}{c c c c c c}
    \hline
    \hline
        
    Galaxy & $N_\mathrm{clouds}$ & $L_\mathrm{MA}$ & $M$ & $\Sigma_\mathrm{avg}$ & AR$_\mathrm{MA}$ \\
    
    &  & (pc) & ($10^3$~M$_\odot$) & ($\mathrm{M}_\odot \mathrm{pc}^{-2}$) & \\
        
    (1) & (2) & (3) & (4) & (5) & (6) \\ 
    
    \hline

    NGC\,4689 & 10,649  & $13_{10}^{20}$ & $4.8_{3.1}^{8.9}$ & $7.7_{7.2}^{8.4}$ & $2.9_{2.3}^{4.2}$ \\
    
    NGC\,628 & 4,467 & $31_{18}^{58}$ & $22_{11}^{57}$ & $7.3_{7.0}^{7.8}$ & $4.1_{2.6}^{6.8}$ \\

    NGC\,1566 & 20,606 & $14_{10}^{22}$ & $6.1_{3.7}^{12.9}$ & $8.7_{8.0}^{9.9}$ &  $3.13_{2.36}^{4.5}$ \\
    
    NGC\,4321 & 24,003 & $13_{9}^{20}$ & $6.2_{3.7}^{13.2}$ & $9.2_{8.1}^{11.3}$ & $2.8_{2.2}^{4}$  \\

    \hline
    
    Centre/Bar & 3,258 & $14_{10}^{26}$ & $10.2_{5.1}^{29.2}$ & $11.3_{8.7}^{17.6}$ &  $2.9_{2.2}^{4.2}$ \\
    
    Disc & 56,467  & $14_{10}^{23}$ & $6.2_{3.6}^{13.5}$ & $8.5_{7.7}^{9.9}$ &  $3_{2.3}^{4.4}$ \\
    
    \hline
    
    Spiral arms & 13,845 & $14_{10}^{24}$ & $7.6_{4.1}^{18.6}$ & $9.2_{8}^{11.7}$ & $2.9_{2.3}^{4.3}$  \\
    
    Inter-arm & 32,103 & $14_{10}^{23}$ & $6.2_{3.7}^{13.6}$ & $8.6_{7.8}^{9.8}$ &  $3_{2.3}^{4.5}$ \\
    
    \hline

    \end{tabular}
    \caption{Characteristics of the homogenised cloud sample. (1) Galaxy name (or selected environment). (2) Number of clouds per environment, $N_\mathrm{clouds}$. (3) Cloud length from the medial axis, $L_\mathrm{MA}$. (4) Cloud mass, $M$. (5) Average gas surface density of clouds, $\Sigma_\mathrm{avg}$. (6) Medial axis aspect ratio, $\mathrm{AR}_\mathrm{MA}$. For columns (3)-(6), the median of the distribution is presented, with the 25$^\mathrm{th}$ and 75$^\mathrm{th}$ percentiles being the subscript and superscript, respectively. The centre/bar, disc (spiral arms + inter-arm), spiral arms and inter-arm populations (bottom 4 rows) show the relevant values for the total cloud sample within the denoted environments. The spiral arms and inter-arm populations do not include clouds from NGC\,4689, since it is a flocculent galaxy.}
    \label{tab:uniform_sample}
\end{table}

\begin{figure}
    \centering
    \includegraphics[width=0.4\textwidth]{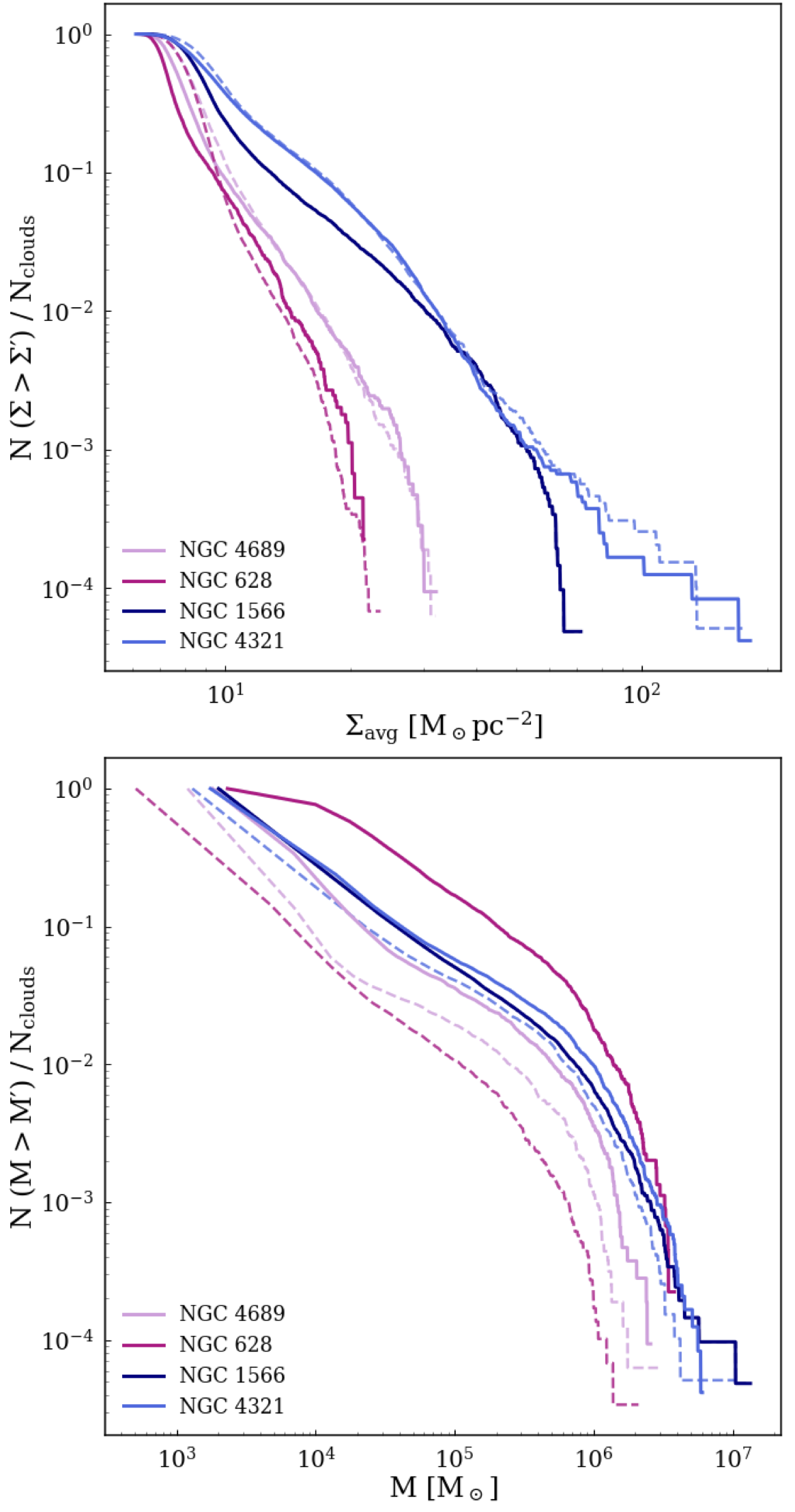}
    \caption{Global cumulative distributions for the average cloud surface density ($\Sigma_\mathrm{avg}$, \textit{top}) and cloud mass ($M$, \textit{bottom}), across the sample (NGC\,4689 in light pink, NGC\,628 in dark pink, NGC\,1566 in dark blue, and NGC\,4321 in light blue). The solid lines show the relevant distributions at the common, homogenised resolution, whilst the dashed lines show the distribution at the relevant galaxy's native resolution. All distributions are normalised by the total number of clouds for that distribution, $N_\mathrm{clouds}$ (as listed in Table~\ref{tab:uniform_sample}).}
    \label{fig:global_CMDs_uni}
\end{figure}

\section{Cloud trends across galaxies}
\label{sec:uni_cat}

To draw any conclusions from the comparison of different galaxies, and the environments within them, it is important to make this comparison as fair as possible. In this Section, we analyse a homogenised or common resolution catalogue built from the surface density maps of each galaxy convolved to match the physical resolution of the furthest target in the sample (i.e. NGC\,1566, see Section~\ref{sec:cloud_extraction}). The resulting composition of the combined catalogue, as well as the statistics of the cloud properties studied here, are listed in Table~\ref{tab:uniform_sample}. In this Section, we explore any correlations between large-scale environments and the characteristics of clouds in the homogenised sample. In particular, special attention is paid to any differences between the centre/bar versus disc clouds, as well as arm versus inter-arm. Additionally, these observed trends at a common resolution are compared to those seen in the previous Section (i.e. at the native physical scale). This is done to study the impact of homogenising multi-galaxy cloud catalogues, as well as performing a qualitative environmental analysis, rather than a more in-depth galaxy-per-galaxy study. 

\subsection{Effect of homogenised resolution on observed trends}

\begin{figure*}
    \centering
    \includegraphics[width=\textwidth]{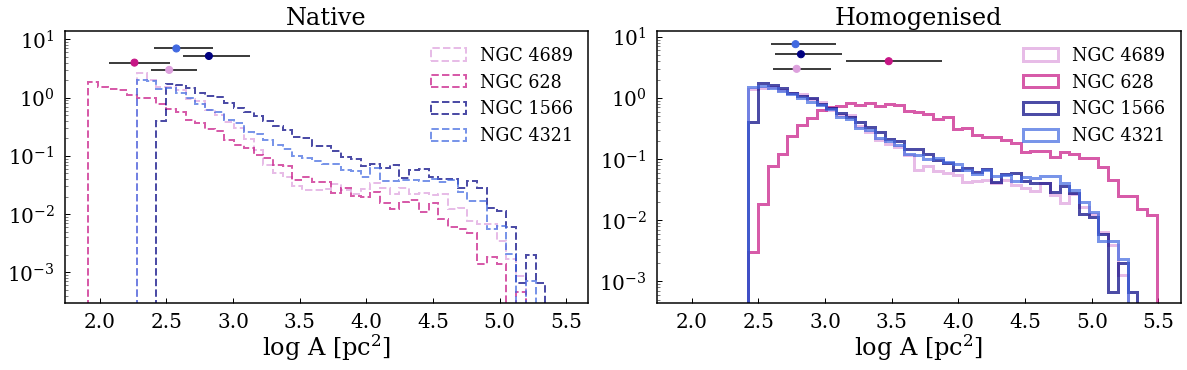}
    \caption{Histograms of cloud area ($A$) for both the native resolution (\textit{left}) and homogenised resolution (\textit{right}), across the homogenised sample (NGC\,4689 in light pink, NGC\,628 in dark pink, NGC\,1566 in dark blue and NGC\,4321 in light blue). Above each distribution are the respective median (solid circle in relevant colour) and interquartile range (solid black line). The histograms are normalised with the respective number of clouds.}
    \label{fig:area_uni_nat}
\end{figure*}

Figure~\ref{fig:global_CMDs_uni} showcases the global cumulative distributions of both cloud surface density and mass for the homogenised sample, as well as the native resolution catalogues. Overall, the observed distributions of $\Sigma_\mathrm{avg}$ appear similar in the native and homogenised catalogues. There are, however, some differences at low values, particularly for NGC\,628. The distributions of cloud masses, on the other hand, see significant changes between the cloud catalogues extracted at different resolutions. For NGC\,4689 and NGC\,628, the cloud mass distributions at the native resolution display consistently lower mass values than the homogenised distributions. Although less pronounced, this is also seen for the barred galaxy NGC\,4321. Given that the observed distributions of cloud surface densities are similar at both resolutions, it is likely that, on average, cloud sizes have increased in the homogenised sample for these galaxies, relative to the native resolution catalogue. 

Figure~\ref{fig:area_uni_nat} displays the distribution of cloud areas at each galaxy's native physical resolution, as well as for the homogenised resolution. As was already hinted with Fig.~\ref{fig:global_CMDs_uni}, for each galaxy except NGC\,1566 (whose physical resolution was defined as the common resolution), clouds are typically larger in the homogenised catalogue. For NGC\,4689 and NGC\,4321 this change in cloud sizes is small, but for NGC\,628, the native and homogenised distributions appear drastically different, with the homogenised clouds being a factor $\sim1$~dex larger. 

These discrepancies are likely an effect akin to "beam smearing". In the present sample of galaxies, NGC\,628 is the closest target ($\sim5$~Mpc), whilst the remaining galaxies are approximately at the same distance ($\sim8$--9~Mpc, see Table~\ref{tab:gal_properties}). This means that the physical resolution achieved for NGC\,4689, NGC\,1566 and NGC\,4321 is poorer than NGC\,628, but similar to each other. When we homogenise all galaxies to the same resolution for comparative purposes, little convolution is needed for the bulk of the sample as they are at similar distances to the furthest target, NGC\,1566 (even though this clearly still has an effect on cloud sizes and masses, see Fig.~\ref{fig:global_CMDs_uni}). For NGC\,628, the convolution kernel needed for degrading to the common resolution is larger, which also lowers the noise threshold. Therefore, it is likely that smaller clouds will be blended into larger but more diffuse associations. This can result in larger clouds with larger masses, but similar or even smaller surface densities.

Figures~\ref{fig:global_CMDs_uni} and \ref{fig:area_uni_nat} highlight the care needed in studies of molecular clouds spanning different galaxies, as observed trends of properties such as mass and size are not purely driven by physical processes, but are heavily dependent on resolution (as well as the cloud extraction process itself).

\subsection{Trends between galaxies}

\begin{figure}
    \centering
    \includegraphics[width=0.4\textwidth]{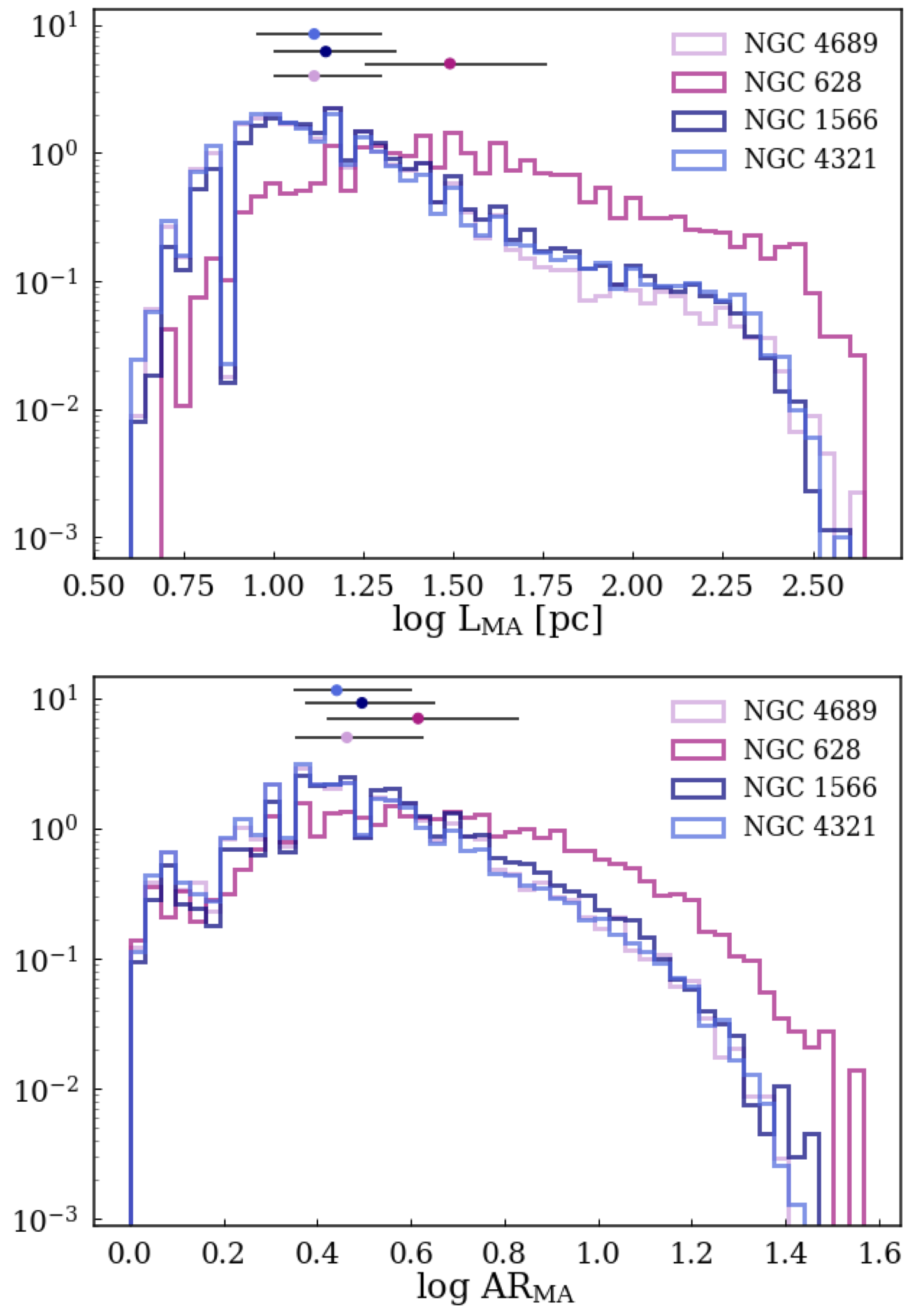}
    \caption{Histograms of cloud length ($L_\mathrm{MA}$, \textit{top}) and cloud aspect ratio ($\mathrm{AR}_\mathrm{MA}$, \textit{bottom}), across the homogenised sample (NGC\,4689 in light pink, NGC\,628 in dark pink, NGC\,1566 in dark blue and NGC\,4321 in light blue). Above each distribution are the respective median (solid circle in relevant colour) and interquartile range (solid black line). The histograms are normalised with the respective number of clouds.}
    \label{fig:Lma_ARma_uni_hist}
\end{figure}

In this Section, we focus on comparing the observed galactic distributions of cloud properties within the homogenised sample. From Fig.~\ref{fig:global_CMDs_uni}, we can see that the barred galaxies (NGC\,1566 and NGC\,4321) tend to have the higher surface density clouds of the sample. The non-barred targets, NGC\,4689 and NGC\,628, display similar $\Sigma_\mathrm{avg}$ distributions, although NGC\,628's distribution seems to fall short in relation to the flocculent's. The two barred targets also display similar cloud mass distributions, which are consistently higher than NGC\,4689. As was already addressed above, NGC\,628 displays high cloud mass values as a likely result of beam smearing.

Figure~\ref{fig:Lma_ARma_uni_hist} showcases the distribution of cloud length and aspect ratio (from the medial axis metric) for the homogenised catalogues of each galaxy. As was already hinted at with Fig.~\ref{fig:global_CMDs_uni}, clouds in NGC\,628 are typically larger (both in $L_\mathrm{MA}$ and area) than the other galaxies in the sample. Cloud aspect ratio values are also slightly higher for this galaxy. The remaining galaxies display similar distributions for both cloud properties, with clouds presenting an average length of $\sim13$~pc and aspect ratio of $\sim3$ (see Table~\ref{tab:uniform_sample}).

In terms of cloud morphology, the distribution of RJ classifications in the homogenised catalogue does not differ from those seen in the native resolution cloud populations (see Section~\ref{sec:rj_nat}). Roughly 33\% of clouds appear circular, 29\% are ring-like, and the remaining 38\% are elongated (16\% centrally overdense and 22\% centrally underdense). Thus, as was also seen in Section~\ref{sec:nat_morphology}, clouds in the homogenised catalogue tend to be quasi-circular (either circular or ring-like). There are no detected variations within different galactic environments.

\begin{figure*}
    \centering
    \includegraphics[width=0.6\linewidth]{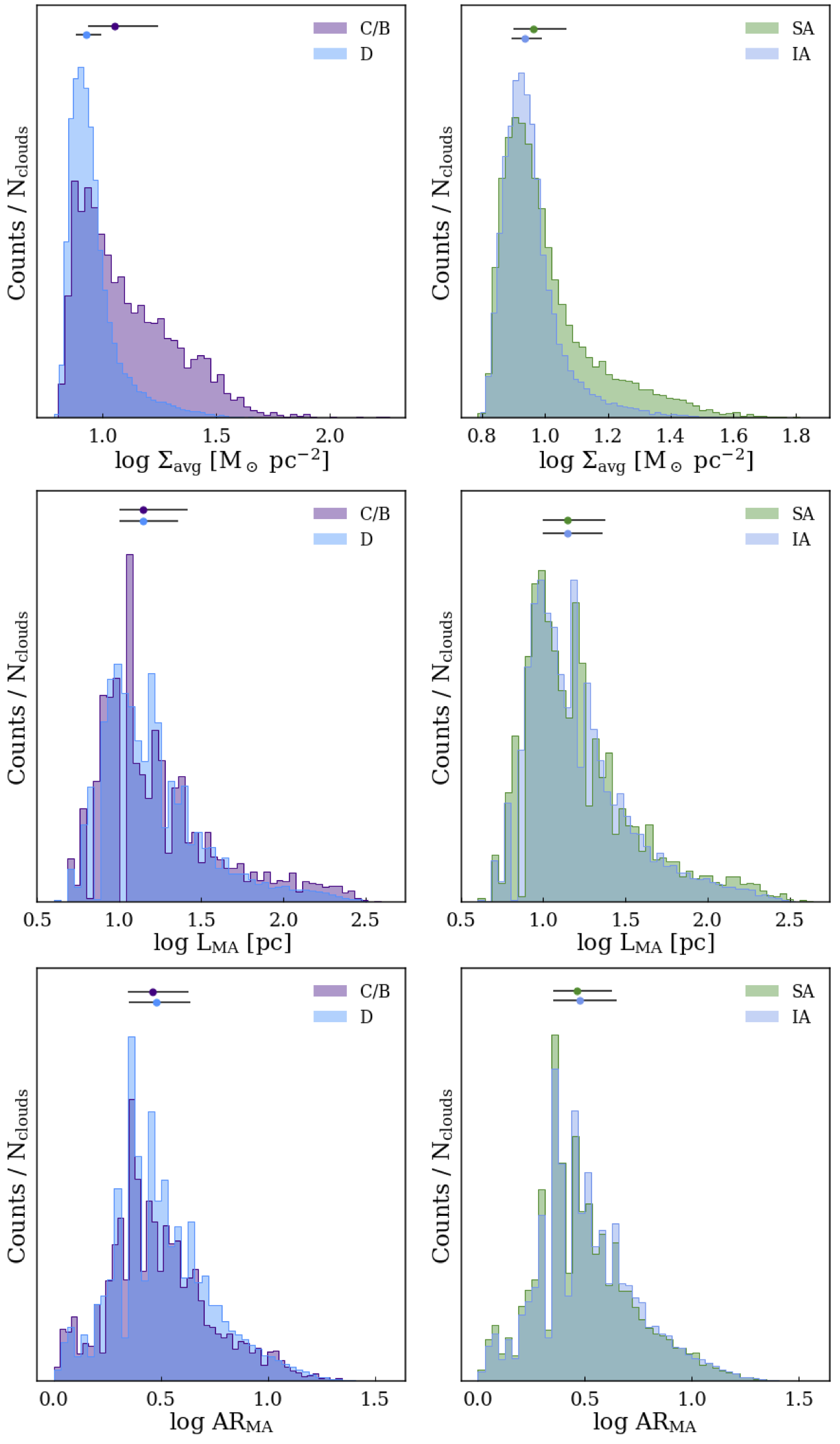}
    \caption{Histograms of the average cloud surface density ($\Sigma_\mathrm{avg}$, \textit{top}), medial axis length ($\mathrm{L}_\mathrm{MA}$, \textit{middle}), and cloud aspect ratio ($\mathrm{AR}_\mathrm{MA}$, \textit{bottom}) for the centre/bar (C/B, purple) versus disc (D, blue) on the left panel, and spiral arms (SA, green) versus inter-arm (IA, light blue) on the right panel, for the combined catalogue of clouds. The y-axis holds the amount of clouds in each bin, normalised by the total number of clouds for the relevant environment ($N_\mathrm{clouds}$). Above each distribution are the respective median (solid circle in relevant colour) and interquartile range (solid black line). The right panel does not include clouds from the flocculent NGC\,4689.}
    \label{fig:split_hist_cd_saia}
\end{figure*}

\subsection{Trends with galactic environment}

In this Section, we focus on analysing cloud properties as a function of large-scale environment across the homogenised cloud catalogue. We particularly focus on the differences (or lack thereof) between the centre/bar and disc clouds, as well as arm versus inter-arm. This type of analysis - where clouds within the same qualitative environment across different galaxies are binned together - is common with large sample studies \citep[e.g.][]{rosolowsky_2021,querejeta_2021,querejeta_2024} when inferring the impact of large-scale environment.

\subsubsection{Centre/Bar versus Disc}
\label{sec:uni_CvsD}

From Table~\ref{tab:uniform_sample}, it is possible to see that the mass and $\Sigma_\mathrm{avg}$ median values are higher in the centre/bar than the disc across the sample. Figure~\ref{fig:split_hist_cd_saia} highlights the differences between the cloud populations in the centre/bar versus the disc. It is possible to see that the distribution of cloud surface densities for the centre/bar has a much larger interquartile spread versus the disc, which is more compact and concentrated towards low-$\Sigma_\mathrm{avg}$ values. This is also observed with cloud mass, although less pronounced. These trends are expected, given that each of the individual galaxies in the sample displayed this trend in the native resolution analysis (Section~\ref{sec:nat_sd}). However, it is important to note that Fig.~\ref{fig:split_hist_cd_saia} portrays only the combined, average trend across the sample. Although we did observe higher surface densities in the centres/bar of all galaxies versus the discs, there are interesting variations observed in our native resolution, galaxy-per-galaxy approach, that would otherwise be missed in this type of combined analysis. For example, from Fig.~\ref{fig:split_hist_cd_saia} it is not possible to tell that this centre/disc disparity is less pronounced in NGC\,628 (Fig.~\ref{fig:splitvp_native_sd}).

Although there are some dissimilarities of cloud sizes and aspect ratio amongst the different galaxies in the homogenised sample (discussed above, see Fig.~\ref{fig:Lma_ARma_uni_hist}), there are no observed differences in these properties when looking at the centre/bar and disc populations of the whole sample. This is clearly shown in Fig.~\ref{fig:split_hist_cd_saia}.

\subsubsection{Spiral arms versus Inter-arm}
\label{sec:uni_SAvsIA}

In the homogenised resolution catalogue, clouds in a spiral arm environment present marginally different average surface densities to clouds in the inter-arm, although this contrast is not as large as what is seen between the centre/bar and the disc. In fact, from Fig.~\ref{fig:split_hist_cd_saia}, it seems that the bulk of both populations is quite similar, with differences arising in the tail-end of the distribution, where there are more high-$\Sigma_\mathrm{avg}$ clouds in the arms. However, this generalised result that clouds on average have higher surface densities in spiral arms than in inter-arm regions can be misleading. Indeed, the behaviour seen in Fig.~\ref{fig:split_hist_cd_saia} is being driven by the more massive barred galaxies, NGC\,4321 and (particularly) NGC\,1566, given that we did not see any discernable differences between the arm and inter-arm environments of NGC\,628 (see Fig.~\ref{fig:splitvp_native_sd}). The relative difference between the arm and inter-arm cloud populations is also different amongst the barred galaxies (see Fig.~\ref{fig:CMD_native}). A generalised environment analysis such as that presented in this Section would have missed these variations which are clues into how different large-scale environments affect the ISM.

Figure~\ref{fig:split_hist_cd_saia} also shows the distribution of $L_\mathrm{MA}$ and $\mathrm{AR}_\mathrm{MA}$ for the cloud populations in the combined spiral arm environments across the sample, as well as the inter-arm. As can be seen from the figure, clouds do not appear to have different sizes or aspect ratio depending on the large-scale environment. This is consistent with what was reported in Section~\ref{sec:nat_morphology} (see also Fig.~\ref{fig:splitvp_native_L_AR}).


\section{Summary and conclusions}
\label{sec:sum_conc}


This paper presents the application of the extinction imaging technique introduced in \cite{faustinovieira_2023} to a wider sample of nearby disc galaxies. This sample consists of a flocculent galaxy (NGC\,4689), a non-barred spiral galaxy (NGC\,628), and two spiral galaxies with bars (NGC\,1566 and NGC\,4321). The goal of this work is to investigate any links between ISM properties and large-scale environment \citep[see also][]{faustinovieira_2024}, for different spiral galaxy morphologies. We also describe the improvements added to the imaging technique to increase its robustness and effectiveness in retrieving extinction features (see Appendix~\ref{app:improvements}). We summarise our findings below:

\begin{itemize}
    \item There is a substantial difference in cloud masses and average surface densities between the centres/bars and discs across the whole sample, with centres/bars showing consistently higher values. The spiral galaxies NGC\,628, NGC\,1566, and NGC\,4321 all show spiral arms with slightly higher medians and extent at smaller galactocentric radii relative to the inter-arm regions. Past a certain $\mathrm{R}_\mathrm{gal}$, both arm and inter-arm distributions become similar. The point at which these converge varies from galaxy to galaxy, from just $\sim$3~kpc for NGC\,628 (which is the fainter spiral), to 8~kpc for NGC\,1566 (which has the smaller bar), and 11~kpc for NGC\,4321 (larger bar). 
    
    \item In the barred galaxies of our sample (NGC\,1566 and NGC\,4321), the average cloud surface density distributions for the arm and inter-arm regions are different inside and outside the radius of the bar, with higher values (and spread) inside $\mathrm{R}_\mathrm{bar}$. This trend is much more pronounced in NGC\,4321, which has a more prominent bar. 
    
    \item There are different radial behaviours for the bars in this sample, with NGC\,4321 holding most of its mass in its very centre, whilst NGC\,1566 does so more towards the ends of the bar and the beginning of the spiral arms. This could hint at different dynamics and gas flows driven by the bars.
    
    \item There is no significant trend of cloud length, size, or aspect ratio with galactic environment across the native resolution catalogue, nor on the homogenised sample. Furthermore, clouds in this work tend to be quasi-circular (i.e. either circular or ring-like), with no detected variations between environments. When looking at the very elongated and long (>100~pc) objects for each galaxy, they appear more "ordered" in NGC\,1566 and NGC\,4321, following the orientation of the more prominent spiral arms, than in NGC\,4689 and NGC\,628, where their orientation seems more random. 

    \item Cloud properties such as mass and size show a dependency on the physical resolution of the data, as global trends shift from the native resolution catalogue to the homogenised resolution one (which was particularly evident for our closest target, NGC\,628). This is because small clouds are likely to be blended into larger associations within a larger beam. Cloud surface densities, on the other hand, are less affected. 

    \item The homogenised resolution catalogue, at first glance, re-iterates the general findings from the native resolution analysis. It does, however, only present an averaged, "combined" effect which is not representative of the conditions within each galaxy. This loss of information when merging cloud catalogues of different galaxies into a single one (under a common resolution) will have an impact on inferring the effect of large-scale environments on the ISM. Clearly, this underlying assumption that all spiral arms, bars, or inter-arm regions behave and affect the ISM the same way is not fully correct.

\end{itemize}

This last result from our work demonstrates the dangers of cloud-population analysis where clouds of a given environment are all binned together. Indeed, a generalised analysis potentially leads to a loss of information on interesting galaxy-to-galaxy variations, which may be key to understanding what drives the properties of molecular clouds. Although small, the present sample of galaxies holds galactic environments that, even though are technically within the same environment "label", look visually distinct (e.g. the bars of NGC\,1566 versus NGC\,4321, the arms of NGC\,628 versus NGC\,1566), and that may also have distinct dynamics which could be affecting the ISM differently. Indeed, the observed difference in cloud masses/surface densities between the arms and inter-arm in this common resolution catalogue is driven solely by the barred targets in the sample. The fact that this does not happen in NGC\,628 (which might point towards a "weaker" spiral potential) is lost in this type of analysis. These types of effects will just be aggravated with increased sample sizes. Therefore, it may be more informative to not only look at environmental trends across multiple galaxies, but also look for any correlation with physical quantities such as pressure, shear, and torques \citep[e.g.][]{miyamoto_2014,sun_2020a,sun_2022,querejeta_2016,ruiz-garcia_2024}. This would help infer if any cloud property variations within each galaxy are linked to any specific environmental conditions, rather than just as a function of a qualitative environment classification. Within the FFOGG project (Following the Flow of Gas in Galaxies\footnote{\url{https://ffogg.github.io/}}), numerical work is underway to simulate analogues of the present sample of galaxies, with which a more quantitative environmental analysis can be performed (Duarte-Cabral et al., \textit{in prep}).

Furthermore, the lack of significant environmental trends for cloud sizes, aspect ratios and morphologies seen here, potentially suggests that investigating these properties on cloud catalogues built from data that has no velocity information is perhaps not the most informative. The shape and size of clouds extracted from position-position maps cannot be assumed to be driven by physical processes alone (e.g. nearby stellar feedback or shear), since effects such as line-of-sight superposition and the specific cloud extraction algorithm used are likely to heavily influence these cloud properties.

This study highlights the care needed when performing molecular cloud property studies as a function of galactic environment, as interesting and potentially enlightening information can be easily lost in the large number statistics of multiple galaxy samples.

The high spatial resolution achieved with our HST extinction mapping technique is comparable to the near/mid-infrared imaging capabilities of JWST. In fact, the present sample of galaxies (excluding NGC\,4689), as well as M51 \citep{faustinovieira_2024}, have now all been imaged with JWST at high-resolution by, for example, the PHANGS-JWST project \citep{lee_2023}, and the FEAST program (Feedback in Emerging extragAlactic Star clusTers; GO1783, PI Adamo). Work is underway on performing a direct comparison between our HST extinction surface densities and those from Polycyclic Aromatic Hydrocarbon (PAH) emission (particularly the $7.7~\upmu\mathrm{m}$ feature). This direct comparison between dust extinction and emission can be a powerful probe of the heating conditions in the ISM \citep[e.g.][]{leroy_2023}, offering more insights into dust physics in nearby galaxies \citep[see also][]{thilker_2023}.

\section*{Acknowledgements}



We thank the anonymous referee for their helpful feedback, which improved the manuscript. HFV and ADC acknowledge the support from the Royal Society University Research Fellowship URF/R1/191609. MWLS and TAD acknowledge the support of a
consolidated grant (ST/W000830/1) from the UK Science
and Technology Facilities Council (STFC). DC and ZB gratefully acknowledge the Collaborative Research
Center 1601 (SFB 1601 sub-project B3) funded by the Deutsche Forschungsgemeinschaft (DFG, German Research Foundation) –
500700252. HFV would like to thank Seamus Clarke and Angela Adamo for helpful discussions. Based on observations made with the NASA/ESA Hubble Space Telescope, which is operated by the Association of Universities for Research in Astronomy, Inc. (Program \#10452). DustPedia is a collaborative focused research project supported by the European Union under the Seventh Framework Programme (2007-2013) call (proposal no. 606847). The participating institutions are: Cardiff University, UK; National Observatory of Athens, Greece; Ghent University, Belgium; Université Paris Sud, France; National Institute for Astrophysics, Italy and CEA, France. This work made use of Astropy:\footnote{\url{http://www.astropy.org}} a community-developed core Python package and an ecosystem of tools and resources for astronomy \citep{astropy:2013, astropy:2018, astropy:2022}.

\section*{Data availability}
With this paper, we release the full catalogue off all clouds extracted with \texttt{SCIMES} and the respective cloud masks in \url{https://dx.doi.org/10.11570/25.0036}, as well as in the FFOGG (Following the Flow of Gas in Galaxies) project website ({\url{https://ffogg.github.io/}}).




\bibliographystyle{mnras}
\bibliography{references} 




\appendix

\section{Improvements on HST extinction technique}
\label{app:improvements}

\begin{figure*}
    \centering
    \includegraphics[width=\textwidth]{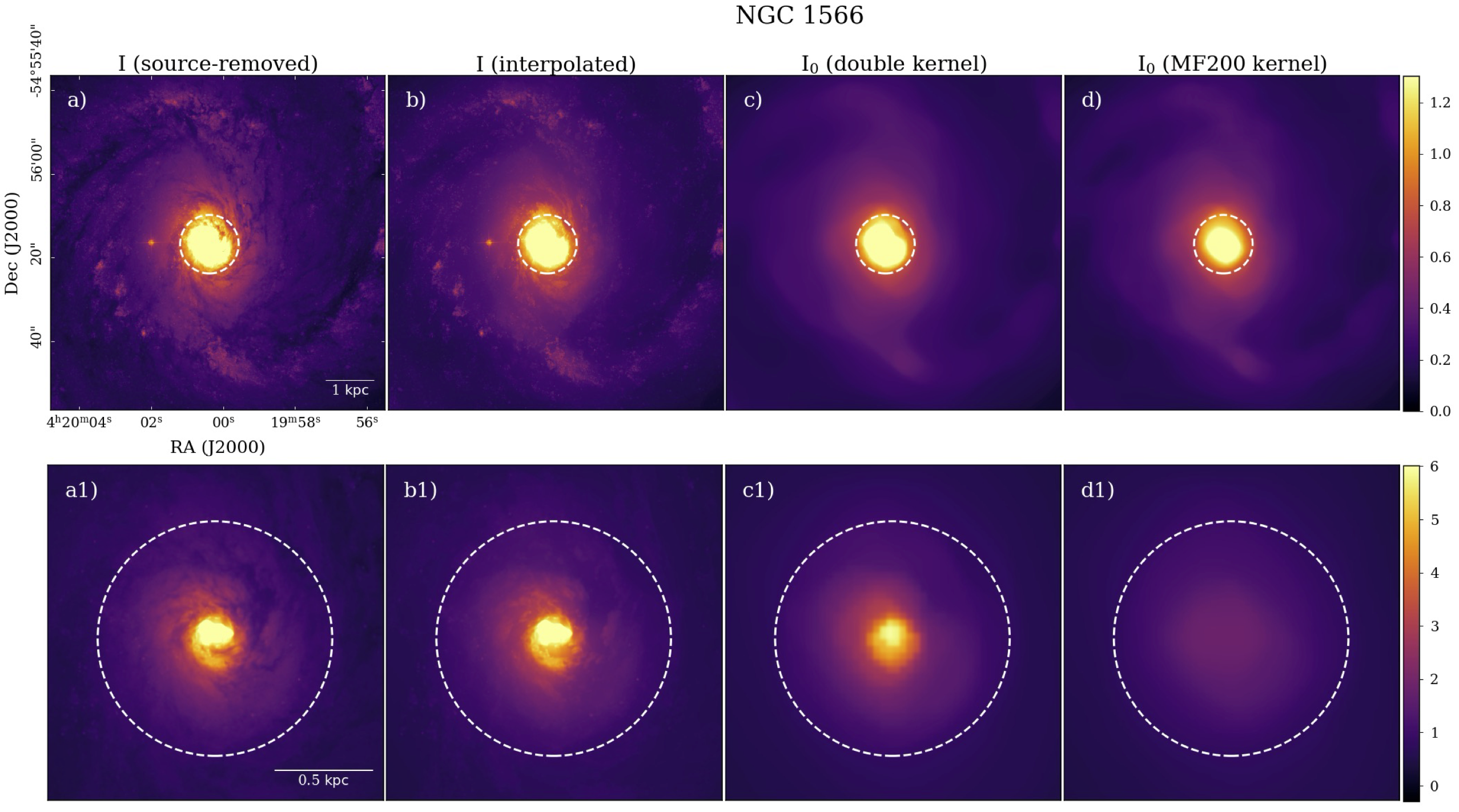}
    \caption{Zoom-in of the central/bar region of NGC\,1566 across different steps in our extinction mapping technique (top row). \textbf{a)} HST V-band observations, $I$ (with bright, point-like sources removed). \textbf{b)} $I$ after linear interpolation of extinction features (as explained in text). \textbf{c)} Reconstructed stellar light map, $I_0$, with two median filters. \textbf{d)} $I_0$ with a single median filter (MF200, $\sim8"$). In the bottom row are the same panels further zoomed-in, and with an adjusted colour scale that better illustrates the differences in the centre. Across all panels, the white circle represents the turnover at which we switch from a smaller kernel in the centre to a larger median filter for the disc (see Fig.~\ref{fig:appA_radial_I0_mf}).}
    \label{fig:app_I_int_I0}
\end{figure*}

\begin{figure}
    \centering
    \includegraphics[width=0.5\textwidth]{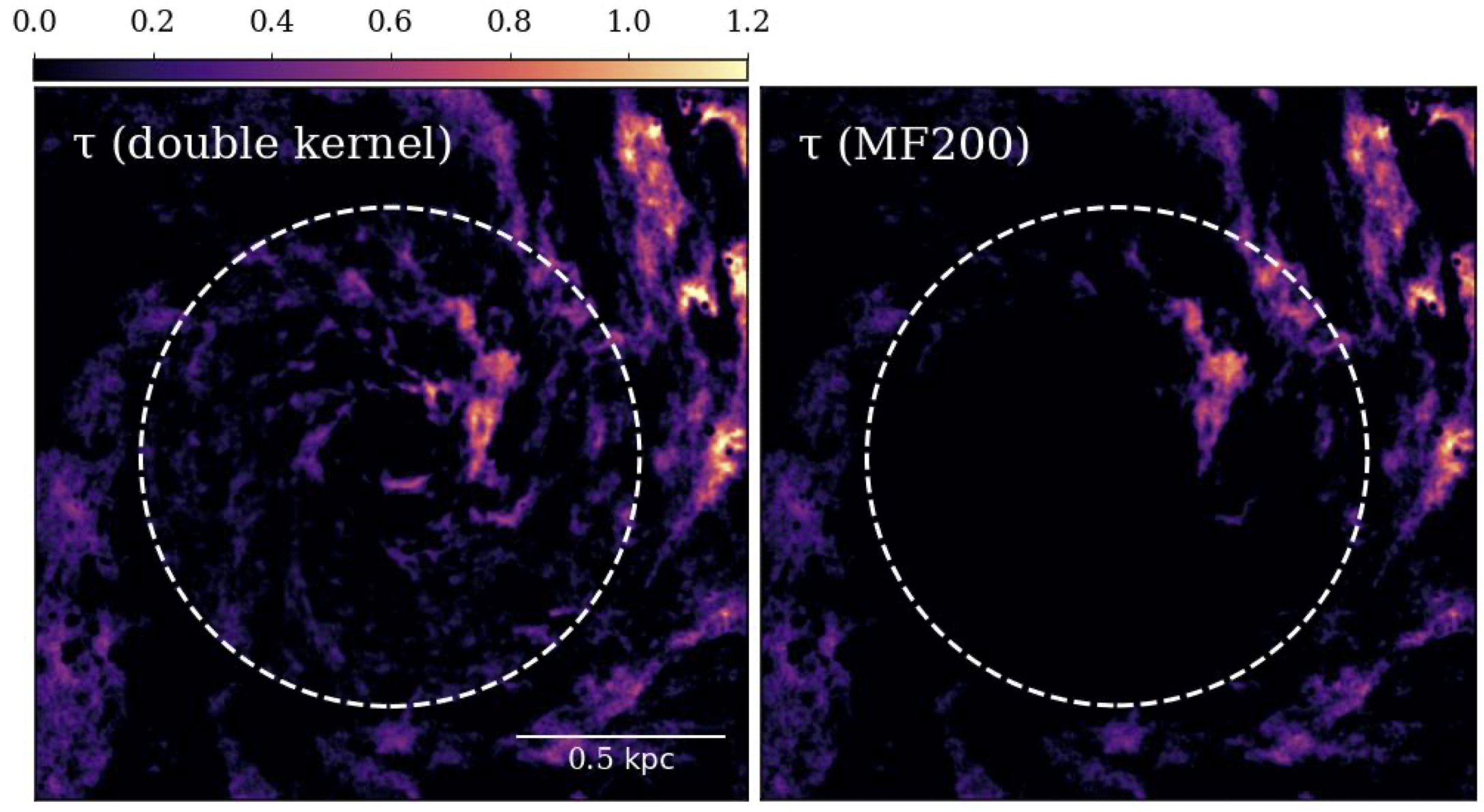}
    \caption{Zoom-in of the central/bar region of NGC\,1566 in the optical depth ($\tau$, Eq.~\ref{eqn:tau_form}) maps for an interpolated, double-kernel stellar light model (left panel), and for a single large median filter $I_0$ (MF200, right panel). Across both panels, the white circle represents the turnover at which we switch from a smaller kernel in the centre to a larger median filter for the disc (see Fig.~\ref{fig:appA_radial_I0_mf}). Note that the colour-scale is the same for both panels.}
    \label{fig:appA_tau}
\end{figure}

This work outlines the application of our high-resolution HST extinction technique \citep{faustinovieira_2023}, with some changes, to 4 nearby galaxies: NGC\,4689, NGC\,628, NGC\,1566 and NGC\,4321. Here, we present the improvements that have been implemented to the technique since \cite{faustinovieira_2023}, as well as our updated pixel-by-pixel uncertainty estimation.

\begin{figure}
    \centering
    \includegraphics[width=0.4\textwidth]{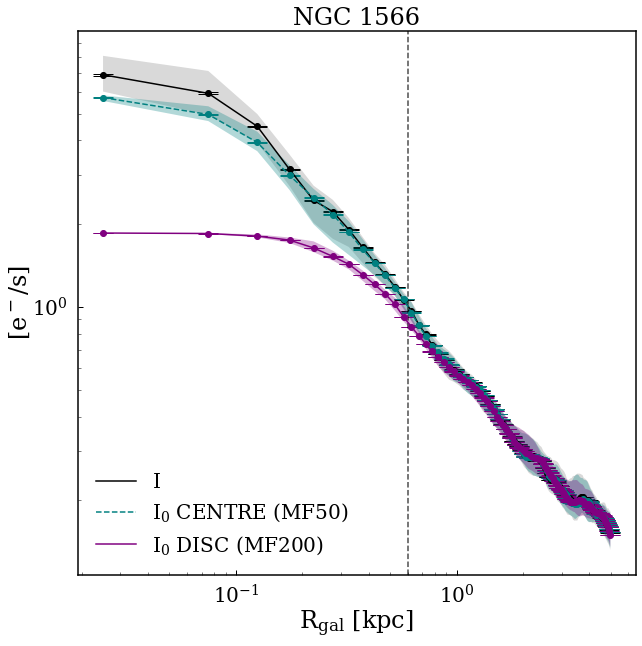}
    \caption{Radial running median profiles for NGC\,1566 of the source-removed HST V-band intensity ($I$, in black), the stellar light model for the centre/bar with a 50\,pix ($\sim2"$) median filter, $I_0$ (in teal), and the disc $I_0$ with a median filter of 200\,pix or $\sim8"$ (in purple). The radial bins have a width of 50\,pc. The shaded region for each profile depicts the respective interquartile range. The errorbars represent the standard error on the median given the bin count, $N$ ($1.253\sigma\sqrt{N}$). The vertical dashed line represents the galactocentric radius at which the larger median filter approximates the $I$ profile.}
    \label{fig:appA_radial_I0_mf}
\end{figure}

\subsection{Stellar light modelling}
\label{app:improvements-Stellar}

Our extinction-based imaging technique estimates dust attenuation by comparing the observed optical light to a smooth, reconstructed stellar distribution. In \cite{faustinovieira_2023}, this stellar distribution ($I_0$) is built by applying a carefully chosen median filter to the HST optical image (after the removal of bright point-like sources). Although careful tests were performed and the final opacity/surface density maps do not change much when varying the chosen median filter by a factor 2, the choice of a kernel to build $I_0$ still remained a human choice.

In this work, we implement some changes such that the choice of kernel minimally impacts the final map, making the technique more robust. In particular, we wanted to improve the estimate of stellar light in highly extinct areas, since large and really prominent dark dust lanes may "contaminate" neighbouring regions when applying a large median filter, resulting in artificially lower values of $I_0$ next to dust lanes. To prevent this, we implemented a masking of extinction regions prior to the reconstruction of the final stellar distribution, so that the dust lanes may be interpolated over without any bleeding. This was done by obtaining an initial optical depth map (Eq.~\ref{eqn:tau_form}) generated with a rough initial guess of $I_0$ (estimated from the source-removed image using a first-guess median filter, e.g. $\sim8"$ for NGC\,1566), and assuming a foreground/background fraction of 50\%. This allows us to retrieve all pixels where the resulting optical depth is above 0 (i.e. where extinction is occurring). These pixels are masked from the source-removed HST V-band image ($I$), and a piecewise linear interpolation is performed to essentially remove the dust lanes from the original $I$ map, thus avoiding contamination when applying median filters (panel b of Fig.~\ref{fig:app_I_int_I0}).

The next step in the method is to create a rough model of the stellar light of each galaxy, using the interpolated source-removed $I$ map. Here, we chose to use a double-component median filter (rather than a single kernel as in \citealt{faustinovieira_2023}), to better retrieve the intensity of the very centre of galaxies, which is more bulge-like. As can be seen from Fig.~\ref{fig:app_I_int_I0}, as well as Fig.~\ref{fig:appA_radial_I0_mf}, if we adopt a single, larger median filter kernel (e.g. $\sim8"$, corresponding to 200 pixels; MF200) for NGC\,1566 for example, we retrieve a reasonable estimate for the disc of the galaxy, but severely underestimate the intensity in the centre of the galaxy (panel d1 of Fig.~\ref{fig:app_I_int_I0}). When the intensity is greatly underestimated, it can lead to the loss of existing extinction features, as is exemplified in Fig.~\ref{fig:appA_tau}. We do not, however, want to adopt a smaller kernel everywhere in the galaxy, as this would not allow to smooth over the larger dust lanes seen in spiral arms, for example. As such, we adopt a double-component approach, where we choose a smaller kernel for the centre, and a larger kernel for the disc. The turnover from the centre to the disc kernel is selected from the radial profile of the maps (shown in Fig.~\ref{fig:appA_radial_I0_mf}), as the point from which the larger disc kernel starts reliably tracing the source-removed V-band profile. As a final step, we smooth the transition from the centre to the disc with a small Gaussian convolution ($\mathrm{FWHM}\sim20$~pix, corresponding to $0.8^{\prime\prime}$ or 38--70~pc). Figure~\ref{fig:app_I_int_I0} showcases what this double-component $I_0$ map looks like against the original HST V-band image (treated to remove point-like sources) as well as the single-component $I_0$ distribution. 

For the centres of our galaxies, we opted to keep our median filters in the 50--100~pix range ($\sim2"\,$--$\,4"$ or $\sim150$\,--$\,290~\mathrm{pc}$), as we found this range more faithfully reproduced the radial profile of the original HST intensity (as exemplified in Fig.~\ref{fig:appA_radial_I0_mf}), whilst still smoothing over any existing extinction features. Likewise, we found that median filters between 200--300~pix ($\sim8"\,$--$\,12"$ or $\sim570$\,--$\,720~\mathrm{pc}$) were reasonable approximations of the stellar light in the discs of our galaxies. Varying the chosen median filters by a factor 2, changes the resulting opacity map (prior to calibration) by only $20\%$.

\subsection{FIR SED modelling: PPMAP}
\label{app:improvements-SED}

\begin{figure*}
    \centering
    \includegraphics[width=0.83\textwidth]{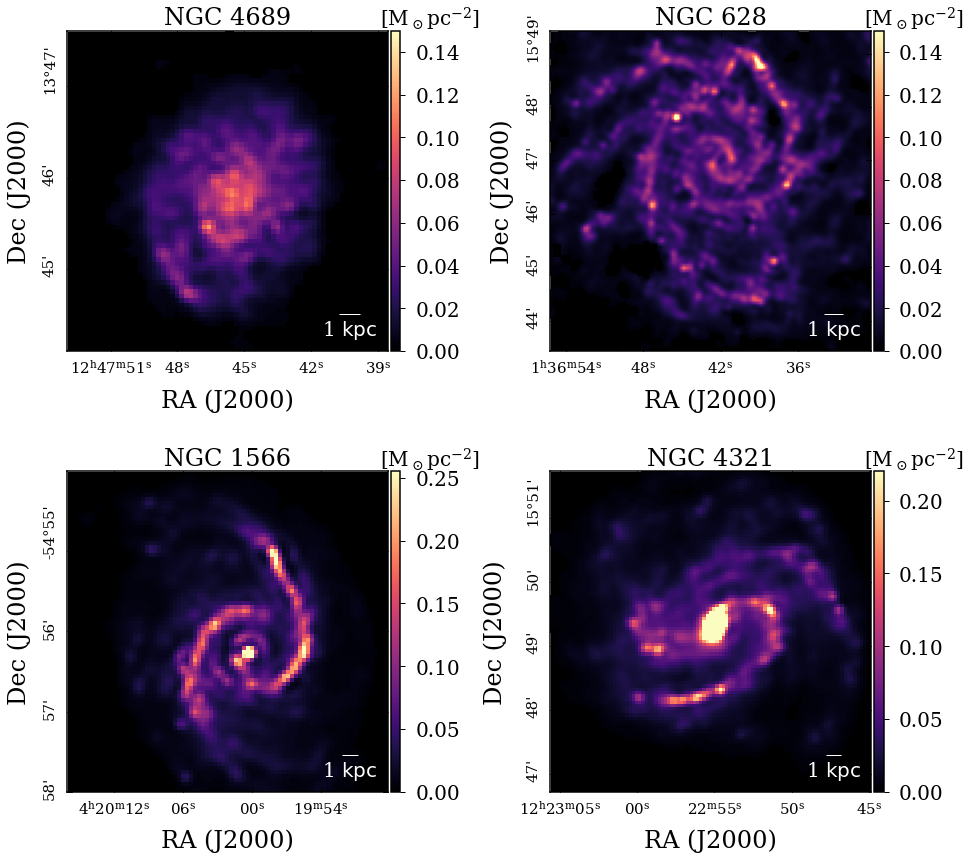}
    \caption{Dust surface density maps resulting from the PPMAP fitting of FIR dust emission observations for NGC\,4689, NGC\,628, NGC\,1566 and NGC\,4321.}
    \label{fig:ppmap_sd}
\end{figure*}

\begin{figure*}
    \centering
    \includegraphics[width=0.83\textwidth]{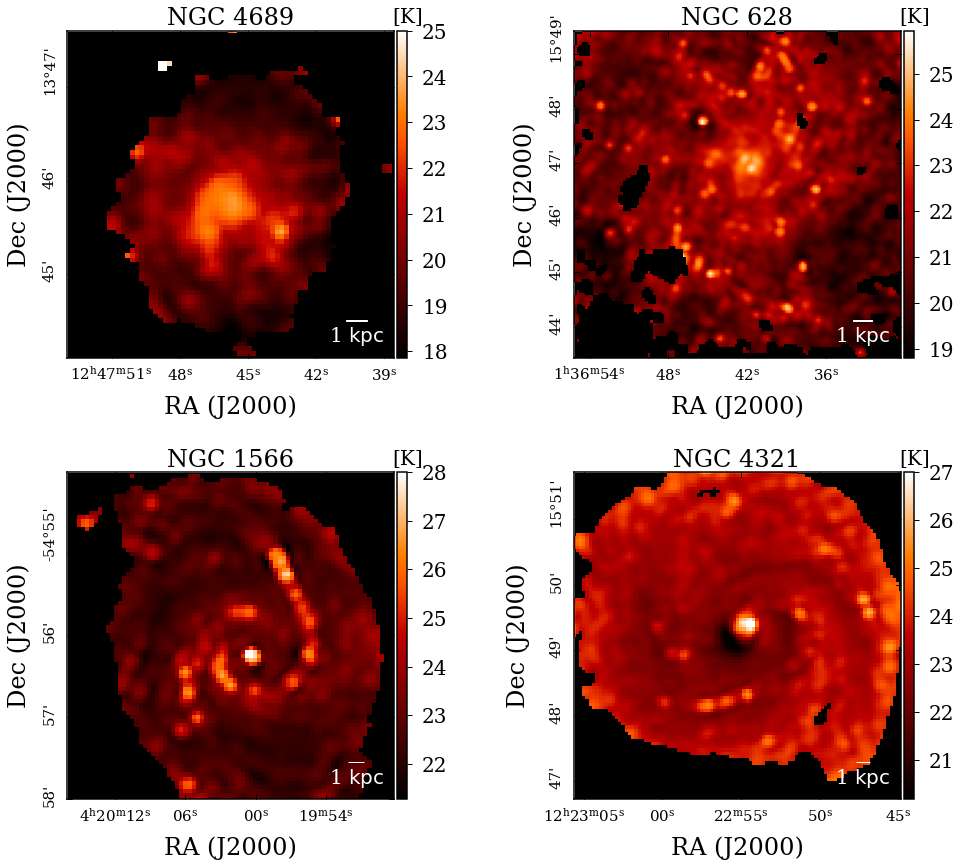}
    \caption{Dust temperature maps resulting from the PPMAP fitting of FIR dust emission observations for NGC\,4689, NGC\,628, NGC\,1566 and NGC\,4321.}
    \label{fig:ppmap_T}
\end{figure*}

In \cite{faustinovieira_2023,faustinovieira_2024}, FIR dust masses (which are used as a benchmark) were derived through simple, single-temperature modified blackbody fits to \textit{Herschel} data. Here, PPMAP \citep[Point Process Mapping;][]{marsh_ppmap_2015} is used instead to create higher resolution dust images of each galaxy in the sample. PPMAP is a Bayesian procedure which fits multi-wavelength data with blackbodies, for a set grid of temperatures. In summary, PPMAP generates a map with increased noise and subsequently decreases this noise in a stepwise approach until it reaches the uncertainty of the input data. With each step, PPMAP minimises the reduced $\chi^2$ between the observed and model values at each pixel, until it finds the optimal solution for both dust temperature and mass at each pixel \citep[for more details see][]{marsh_ppmap_2015}. Unlike traditional spectral energy distribution (SED) modelling, PPMAP uses the relevant instrument's point-spread function (PSF) into account, and therefore does not require the images to be degraded to a common resolution. This allows PPMAP to provide higher-resolution dust maps than standard SED fitting techniques, which are approximately limited by the highest resolution image provided ($\sim5"$ for $70\,\upmu\mathrm{m}$ for our galaxies, except for NGC\,4689 where it is $\sim7"$). To run PPMAP, we assume a fixed dust emissivity index ($\beta$) of 1.8 \citep{planck_2011}, and a dust mass absorption coefficient of $\kappa_{250\mu\mathrm{m}}=21.6~\mathrm{cm}^2\mathrm{g}^{-1}$ \citep[from][]{ossenkopf_dust_1994}. For the PPMAP fits, we initially used 13 values of temperature, logarithmically spaced between 10--80\,K. However, we found that in the very lowest and highest temperature bins, no significant dust column was found. Since including these temperature bins also created large uncertainties (most likely due to limited wavelength coverage), we removed these temperature slices and restricted the fits to only consider temperatures between 16.8--40\,K (or 16.8--67\,K for NGC\,628, since for this galaxy, there were some significant contributions from hot dust detected). The final dust emission surface density map (and temperature map) was created with the results from the PPMAP fits (i.e. summing through each temperature slice), and we use only the pixels whose values are greater than 3$\sigma$ \footnote{The standard deviation $\sigma$ is modelled by PPMAP, considering the photometric noise of each input image.} (shown in Fig.~\ref{fig:ppmap_sd}). Similarly, the temperature maps for each galaxy are an average across all temperature slices, weighted by the measured dust surface density at each bin (Fig.~\ref{fig:ppmap_T}).

\subsection{Monte Carlo uncertainties}
\label{appA:monte_carlo}

\begin{table}
    \centering
    \begin{tabular}{c|c c c}
        \hline
        \hline
        Galaxy & log $\sigma_I$ & log $\sigma_{f/b}$ & log $\sigma_{I_0}^*$ \\ 
         
        \hline

        NGC\,4689 & -2.28 & -2.71 & -2.122 (-2.125) \\
        NGC\,628 & -2.18 & -1.5 & -2.025 (-2.026) \\
        NGC\,1566 & -1.94 & -2.68 & -1.791 (-1.792) \\
        NGC\,4321 & -1.89 & -2.07 & -1.737 (-1.739) \\
        
        \hline

        \multicolumn{4}{l}{$^{*}$ Values shown for centre/bar (disc), given the different median filters.} \\

    \end{tabular}
    \caption{The uncertainties (in log-form) of the quantities used to estimate the relative error of the final opacity (and surface density) map for each galaxy, on a pixel-by-pixel basis. $\sigma_I$ is the photometric noise of the V-band image, after point-like source removal. $\sigma_{f/b}$ is the uncertainty on the calibration with dust emission, which we take to be the scatter on the measured foreground/background fraction (see text). $\sigma_{I_0}$ is the uncertainty on the reconstructed stellar light map.}
    \label{tab:MC_unc}
\end{table}

To measure the uncertainty in our opacity estimates, $\tau$, we perform Monte Carlo realizations for each pixel in our maps. As explained in \cite{faustinovieira_2023}, $\tau$ will depend on the photometric noise of the HST V-band image, $\sigma_I$, the associated error when constructing the stellar light models, $\sigma_{I_0}$, and the uncertainty on the calibration with FIR dust emission observations, which we take to be the scatter on the foreground/background fraction resulting from said calibration, $\sigma_{f/b}$.

Since the methodology on the stellar light modelling was changed from that of \cite{faustinovieira_2023} (as explained in Appendix\,\ref{app:improvements-Stellar}), the calculation of $\sigma_{I_0}$ is also updated accordingly. Before, with a single median filter, we computed the uncertainty through the standard error on the median, $\sigma_\mathrm{MF} = 1.2533\,\sigma_I\,\sqrt{\mathrm{MF}}$, where MF is the respective median filter in pixels. Here, this calculation is performed for the two median filters used, and additionally, the uncertainty on the linear interpolation used when building these maps is also accounted for. Considering that the uncertainty in a linear interpolation ($\sigma_\mathrm{int}$) can be approximated by propagating the uncertainties between 2 points\footnote{This is the approach for a linear interpolation in a 1D case, which serves to give a conservative estimate.}, then

\begin{equation}
    \sigma_\mathrm{int}=\sqrt{\sigma_I^2+\sigma_I^2}\sim\sqrt{2}\,\sigma_I,
\end{equation}

\noindent assuming that all points in the map have uncertainty equal to the photometric noise. Finally, the error on the double-component $I_0$ map can be estimated by propagating $\sigma_\mathrm{MF}$ and $\sigma_\mathrm{int}$, such that

\begin{equation}
    \sigma_{I_0}=\sqrt{\sigma_\mathrm{MF}^2+\sigma_\mathrm{int}^2}.
\end{equation}

Table~\ref{tab:MC_unc} holds the different uncertainties that are used when performing the $10^4$ pixel-by-pixel Monte Carlo simulations for each galaxy. These can be propagated to give the uncertainty on the total dust mass we retrieve for each galaxy in the sample (shown in Table~\ref{tab:gal_properties}). It is important to note that these uncertainties do not take into account any errors associated with the assumed distance to a galaxy, dust mass absorption coefficient, or any other systematic errors.

\section{Cloud catalogues}
\label{app:catalogue}

\begin{table*}
    \begin{tabular}{ l | p{14.5cm} }
    \hline
    \hline
    Catalogue Column & Description \\
    \hline
    \textit{ID}  & Unique ID number of cloud \\
    \textit{Galaxy}$^{*}$ & Host galaxy name \\
    \textit{RA\_deg} & Right ascension of cloud (degrees) \\
    \textit{Dec\_deg}  & Declination of cloud (degrees) \\
    \textit{RA\_pix}  & Right ascension of cloud (pixel coordinates) \\
    \textit{Dec\_pix}  & Declination of cloud (pixel coordinates) \\
    \textit{R\_gal} & Distance of cloud centre to the galactic centre (kpc) \\
    \textit{Sigma\_tot}  & Total sum of the gas mass surface density of every pixel in the cloud ($\mathrm{M}_\odot~\mathrm{pc}^{-2}$) \\
    \textit{Sigma\_avg}  & Average gas mass surface density of cloud ($\mathrm{M}_\odot~\mathrm{pc}^{-2}$) \\
    \textit{Area\_ellipse} & Area of the ellipse defined by the second moments of the cloud (pc$^{2}$) \\
    \textit{Area\_exact}  & Exact area of cloud (pc$^{2}$) \\
    \textit{R\_eq}  & Equivalent radius estimated using the cloud's exact area (i.e. $R_\mathrm{eq} = \sqrt{A/\pi}$; pc) \\
    \textit{Mass}  & Mass of cloud (M$_\odot$) \\
    \textit{Major\_axis\_a}  & Semi-major axis (pc) \\
    \textit{Minor\_axis\_b}  & Semi-minor axis (pc) \\
    \textit{AR\_ab}  & Aspect ratio between semi-major and semi-minor axis \\
    \textit{PA} & Position angle of cloud major axis, measured counter-clockwise from $+x$ axis (degrees) \\
    \textit{Length\_MA}  & Length of the geometrical medial axis (pc) \\
    \textit{Width\_MA}  & Width of the geometrical medial axis (pc) \\
    \textit{AR\_MA}  & Aspect ratio between the medial axis length and width \\
    \textit{RJ1} & Rotated $J$-value, $\mathrm{R}_1$ \\
    \textit{RJ2} & Rotated $J$-value, $\mathrm{R}_2$ \\
    \textit{RJ\_class} & Rotated $J$-value cloud morphology classification. \\ & 
    (1=circular-like, 2=ring-like, 3=centrally overdense filament-like, 4=centrally underdense filament-like) \\
    \textit{Rel\_err} & Relative uncertainty on the cloud's opacity (and thus surface density/mass) from the dust extinction technique alone \\    
    \textit{Env} & Tag identifying the environment of the cloud, as per \cite{querejeta_2021}. \\ & (Centre/Bar=1,2,3. Spiral arms=5,6. Inter-arm=4,7,10) \\
    \textit{Not\_edge\_cut} & Tag identifying clouds that do not touch the edge of the map (1=not edge, 0=edge) \\
    \textit{Rel\_err\_cut} & Tag identifying clouds that pass the relative uncertainty $<30\%$ criteria (1=yes, 0=no) \\
    \textit{Tau\_max\_cut} & Tag identifying clouds that have less than 30\% of their pixels with $\tau>\tau_\mathrm{max}$ (1=yes, 0=no) \\
    \textit{Uncertain\_mass\_tag}$^+$ & Tag identifying clouds with \textit{Tau\_max\_cut}=0, but we choose to keep in the fiducial sample (1=yes, 0=no) \\
    \hline

    \multicolumn{2}{l}{$^{*}$ Only for the homogenised resolution catalogue.} \\
    \multicolumn{2}{l}{$^{+}$ Only for the NGC\,4321 catalogue.}
    \end{tabular}
    \caption{Description of the cloud catalogues obtained in this work.}
    \label{tab:catalogue}
\end{table*}

With this paper, we release the complete cloud masks and their respective catalogues\footnote{\url{https://ffogg.github.io/ffogg.html}} of each galaxy presented in this work (as per Section~\ref{sec:cloud_extraction}), as well as the homogenised resolution catalogue. Table~\ref{tab:catalogue} specifies all the cloud properties of the catalogues.

\subsection{Extreme clouds ($\Sigma_\mathrm{avg}$ and LSFs)}

In Sections~\ref{sec:extreme_sd} and \ref{sec:extreme_lsf}, we analyse the spatial distribution of the highest surface density clouds of each galaxy, as well as clouds that approximate large-scale filaments (elongated, $L_\mathrm{MA}>100$~pc). To determine the statistical significance of these distributions, we perform $10^6$ random draws of a subset of $N$ clouds, and calculate the $\chi^2$ value for each draw. From the cumulative distribution of the simulated $\chi^2$ values, it is possible to determine the likelihood ($\mathrm{p}_\mathrm{rnd}$) of obtaining a given distribution (i.e. $\chi^2$) from random sampling. All of the extreme cloud distributions considered here are statistically significant (i.e. $\mathrm{p}_\mathrm{rnd}<10^{-4}$).

\begin{table*}
    \centering
    \begin{tabular}{c | c c | c c}
    \hline
    \hline

    Galaxy & $N_\Sigma$ & $\mathrm{p}_\mathrm{rnd,\Sigma}$ & $N_\mathrm{LSF}$& $\mathrm{p}_\mathrm{rnd, LSF}$ \\
    (1) & (2) & (3) & (4) & (5) \\
    
    \hline
    
       NGC\,4689 & 667 & -- & 126 & -- \\

       NGC\,628 & 1309 & $<5\times10^{-6}$ & 265 & $<3\times10^{-6}$ \\

       NGC\,1566 & 1030 & $<9\times10^{-6}$ & 309 & $<1\times10^{-6}$ \\

       NGC\,4321 & 832 & $<1\times10^{-6}$ & 272 & $<2\times10^{-6}$ \\
       
    \hline
    \end{tabular}
    \caption{Statistics from the Pearson $\chi^2$ analysis performed for the extreme cloud subsamples. Columns (2)--(3) refer to the highest surface density objects in each galaxy (see Section~\ref{sec:extreme_sd}), whilst columns (4)--(5) are referring to clouds in the catalogues that approximate large-scale filaments (see Section~\ref{sec:extreme_lsf}). (1) Galaxy name. (2) Number of clouds in the high-$\Sigma_\mathrm{avg}$ subsample, $N_\Sigma$. (3) Likelihood of obtaining a given $\chi^2$ from a pure random draw of $N_\Sigma$ clouds. (4) Number of large-scale filaments, $N_\mathrm{LSF}$. (5) Likelihood of obtaining a given $\chi^2$ from a pure random draw of $N_\mathrm{LSF}$ clouds.}
    \label{tab:chi2_sd_lsf}
\end{table*}

\subsection{RJ classification}

\begin{table*}
    \centering
    \begin{tabular}{c|c c c c| c}
         \hline 
         \hline
         Environment & \multicolumn{4}{c|}{RJ-class (\%)} & p-value \\
          & 1 & 2 & 3 & 4 &  \\
          
         \hline
         
         NGC 4689 & 22 & 34 & 17 & 27 & --\\
         Centre & 37 & 23 & 21 & 19 & 0.0015 \\
         Disc & 22 & 34 & 17 & 27 & 0.99 \\

         \hline
         
         NGC 628 & 26 & 34 & 15 & 25 & --\\
         Centre & 33 & 30 & 12 & 25 & 0.41 \\
         Spiral arms & 27 & 34 & 15 & 24 & 0.99 \\
         Inter-arm & 25 & 34 & 15 & 26 & 0.99 \\

         \hline

         NGC 1566 & 27 & 34 & 14 & 25 & --\\
         Centre & 32 & 8 & 28 & 32 & $5.12\times10^{-8}$ \\
         Bar & 37 & 23 & 13 & 26 & 0.051 \\
         Bar ends & 33 & 31 & 14 & 22 & 0.611 \\
         Spiral arms ($<\mathrm{R}_\mathrm{bar}$) & 33 & 31 & 11 & 25 & 0.459 \\
         Inter-bar & 33 & 33 & 13 & 21 & 0.577 \\
         Spiral arms ($>\mathrm{R}_\mathrm{bar}$) & 32 & 30 & 15 & 23 & 0.72 \\
         Inter-arm ($>\mathrm{R}_\mathrm{bar}$) & 25 & 36 & 14 & 25 & 0.98 \\

         \hline

         NGC 4321 & 28 & 33 & 16 & 23 & --\\
         Centre & 42 & 18 & 26 & 13 & $1.25\times10^{-5}$ \\
         Bar & 30 & 31 & 15 & 24  & 0.98 \\
         Bar ends & 37 & 28 & 15 & 20 & 0.27 \\
         Spiral arms ($<\mathrm{R}_\mathrm{bar}$) & 29 & 35 & 14 & 22 & 0.91 \\
         Inter-bar & 29 & 32 & 15 & 24 & 0.97 \\
         Spiral arms ($>\mathrm{R}_\mathrm{bar}$) & 29 & 32 & 17 & 22 & 0.99 \\
         Inter-arm ($>\mathrm{R}_\mathrm{bar}$) & 25 & 34 & 16 & 24 & 0.95 \\

         \hline

    \end{tabular}
    \caption{RJ-class distribution (i.e. cloud morphology) for the 4 galaxies in the native resolution sample, and within their respective environments. Circular clouds have RJ=1, for ring-like clouds RJ=2, overdense elongated clouds have RJ=3, and underdense elongated clouds have RJ=4. The last column holds the associated p-value of the Pearson $\chi^2$ analysis performed.}
    \label{tab:RJ_class}
\end{table*}

\begin{figure*}
    \centering
    \includegraphics[width=0.9\textwidth]{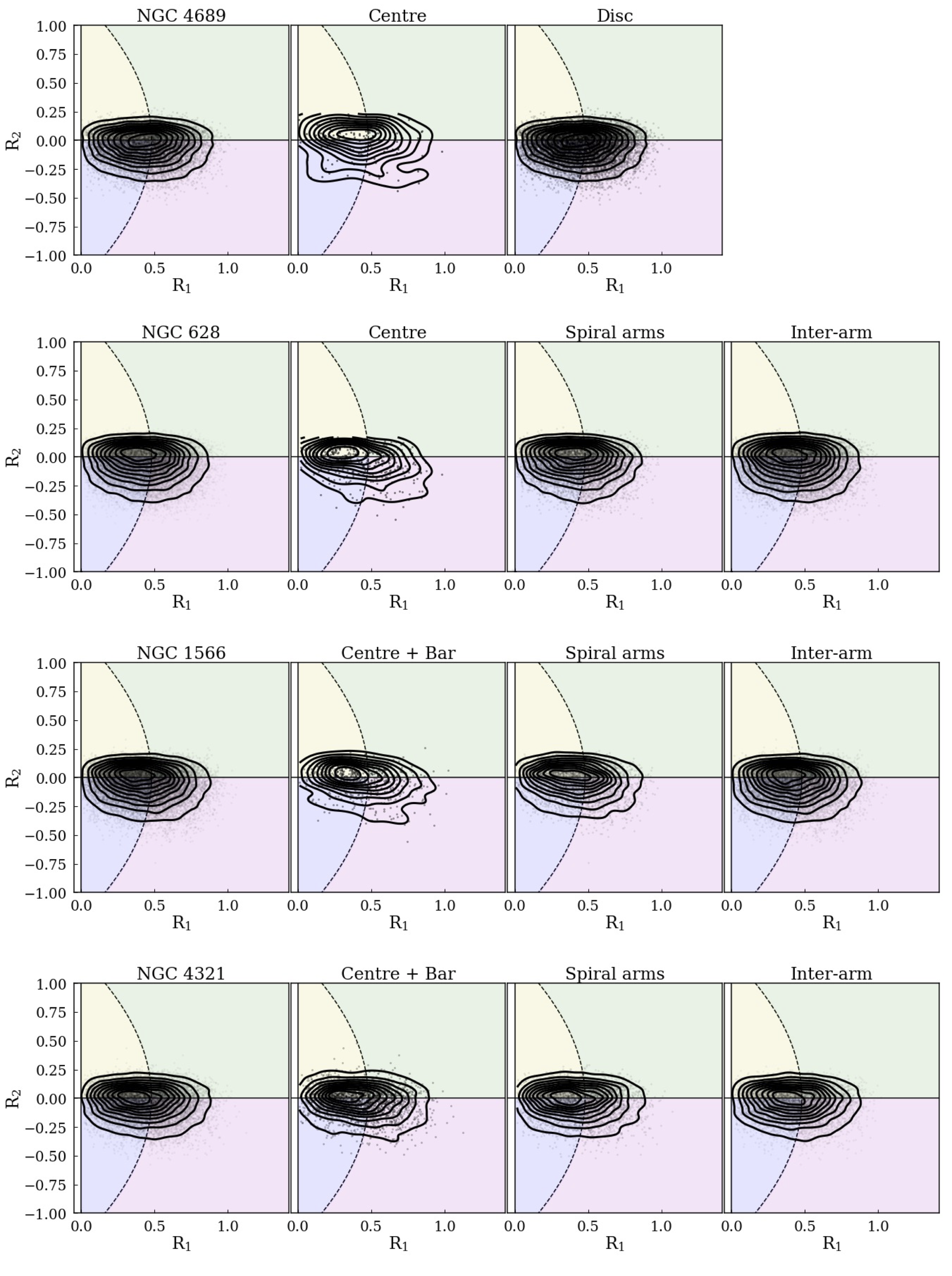}
    \caption{RJ-plots for all galaxies in the sample and their respective environments (native resolution). Clouds are categorised into 4 morphological classes: circular (RJ=1, yellow), ring-like (RJ=2, purple), centrally overdense elongated (RJ=3, green) and centrally underdense elongated (RJ=4, pink).}
    \label{fig:RJ_gal}
\end{figure*}

This Section holds the RJ-plots of all galaxies and their environments for the native resolution catalogue (Fig.~\ref{fig:RJ_gal}), as well as the statistics resulting from the Pearson $\chi^2$ analysis performed (Table~\ref{tab:RJ_class}).

\bsp	
\label{lastpage}
\end{document}